\documentclass[conference]{IEEEtran}
\usepackage{cite}

\ifCLASSINFOpdf
\else
\fi
\usepackage{times}
\usepackage{graphicx} 
\usepackage{subfigure} 

\usepackage{multirow}
\usepackage{mathrsfs}
\usepackage{amssymb}
\usepackage[cmex10]{amsmath}
\usepackage{balance}
\usepackage{cite}
\usepackage[numbers]{natbib}
\usepackage{wrapfig}
\usepackage[bookmarks=false]{hyperref}
\usepackage[noend]{algorithmic}
\usepackage{algorithm}

\usepackage[noend]{algorithmic}

\newtheorem{mydef}{Definition}
\newtheorem{theorem}{Theorem}
\newtheorem{lemma}{Lemma}

\newtheorem{proof}{Proof}

\newcommand{\squishlist}{
 \begin{list}{$\bullet$}
  { \setlength{\itemsep}{0pt}
     \setlength{\parsep}{3pt}
     \setlength{\topsep}{3pt}
     \setlength{\partopsep}{0pt}
     \setlength{\leftmargin}{1.5em}
     \setlength{\labelwidth}{1em}
     \setlength{\labelsep}{0.5em} } }
\newcommand{\squishlisttwo}{
 \begin{list}{$\bullet$}
  { \setlength{\itemsep}{0pt}
     \setlength{\parsep}{0pt}
    \setlength{\topsep}{0pt}
    \setlength{\partopsep}{0pt}
    \setlength{\leftmargin}{2em}
    \setlength{\labelwidth}{1.5em}
    \setlength{\labelsep}{0.5em} } }
\newcommand{\squishend}{
  \end{list}  }

\IEEEoverridecommandlockouts
\begin{document}
%
\title{Adaptive Laplace Mechanism: Differential Privacy Preservation in Deep Learning}

\author{Blind Review}


%
\author{\IEEEauthorblockN{NhatHai Phan\IEEEauthorrefmark{1}$^1$\thanks{$^1$ This is a correction version of the previous arXiv:1709.05750 and ICDM'17 published version. Refer to Appendix G for summary of changes.},
Xintao Wu\IEEEauthorrefmark{2},
Han Hu\IEEEauthorrefmark{1}, 
and Dejing Dou\IEEEauthorrefmark{3}}
\IEEEauthorblockA{\IEEEauthorrefmark{1}New Jersey Institute of Technology, 
Newark, New Jersey, USA}
\IEEEauthorblockA{\IEEEauthorrefmark{2}University of Arkansas, Fayetteville, Arkansas, USA}
\IEEEauthorblockA{\IEEEauthorrefmark{3}University of Oregon, Eugene, Oregon, USA \\ 
Emails: phan@njit.edu, xintaowu@uark.edu, hh255@njit.edu, and dou@cs.uoregon.edu}
}



\maketitle

\begin{abstract}
In this paper, we focus on developing a novel mechanism to preserve differential privacy in deep neural networks, such that: \textbf{(1)} The privacy budget consumption is totally independent of the number of training steps; \textbf{(2)} It has the ability to adaptively inject noise into features based on the contribution of each to the output; and \textbf{(3)} It could be applied in a variety of different deep neural networks. To achieve this, we figure out a way to perturb affine transformations of neurons, and loss functions used in deep neural networks. In addition, our mechanism intentionally adds \textit{``more noise"} into features which are \textit{``less relevant"} to the model output, and vice-versa. Our theoretical analysis further derives the sensitivities and error bounds of our mechanism. Rigorous experiments conducted on MNIST and CIFAR-10 datasets show that our mechanism is highly effective and outperforms existing solutions.
\end{abstract}


%
\IEEEpeerreviewmaketitle

\section{Introduction}
Today, deep learning has become the tool of choice in many areas of engineering, such as autonomous systems, signal and information processing, and data analytics. Deep learning systems are, therefore, not only applied in classic settings, such as speech and handwriting recognition, but also progressively operate at the core of security and privacy critical applications. 

For instance, self-driving cars make use of deep learning for recognizing objects and street signs \cite{Levinson5940562}. Detection systems for email spam integrate learning methods for analyzing data more effectively \cite{Lowd05goodword}. Furthermore, deep learning has applications in a number of healthcare areas, e.g., phenotype extraction and health risk prediction \cite{Feiwang:2016}, prediction of the development of various diseases, including schizophrenia, cancers, diabetes, heart failure, etc. \cite{Choiocw112}, 
and many more. This presents an obvious threat to privacy in new deep learning systems which are being deployed. However, there are only a few scientific studies in preserving privacy in deep learning. 

In the past few decades, a subject of significant interest has been how to release the sensitive results of statistical analyses and data mining, while still protecting privacy. One state-of-the-art privacy model is $\epsilon$-differential privacy \cite{dwork2006calibrating}, which ensures that the adversary cannot infer any information about any specific record with high confidence (controlled by a privacy budget) from the released learning models, even if all the remaining tuples of the sensitive data are possessed by the adversary. The privacy budget controls the amount by which the output distributions induced by two neighboring datasets may differ: A smaller privacy budget value enforces a stronger privacy guarantee. Differential privacy research has been studied from both theoretical and application perspectives \cite{nipsChaudhuriM08,mcsherry2009differentially}.
The mechanisms of achieving differential privacy mainly include adding Laplace noise \cite{dwork2006calibrating}, the exponential mechanism \cite{McSherry:2007}, and the functional perturbation method \cite{nipsChaudhuriM08}. 

It is significant and timely to combine differential privacy and deep learning, i.e., the two state-of-the-art techniques in privacy preserving and machine learning. However, this is a challenging task, and only a few scientific studies have been conducted. In \cite{ShokriVitaly2015}, Shokri and Shmatikov proposed a distributed training method, which injects noise into \textit{``gradients"} of parameters, to preserve privacy in neural networks. In this method, the magnitude of injected noise and the privacy budget $\epsilon$ are accumulated in proportion to the number of training epochs and the number of shared parameters. Thus, it may consume an unnecessarily large portion of the privacy budget, as the number of training epochs and the number of shared parameters among multiple parties are often large \cite{Phan0WD16}. 

To improve this, based on the composition theorem \cite{Dwork:2009:DPR},
Abadi et al. \cite{Abadi} proposed a privacy accountant, which keeps track of privacy spending and enforces applicable privacy policies. However, the approach is still dependent on the number of training epochs, as it introduces noise into \textit{``gradients"} of parameters in every training step. With a small privacy budget $\epsilon$, only a small number of epochs can be used to train the model \cite{Abadi}. In practice, that could potentially affect the model utility, when the number of training epochs needs to be large to guarantee the model accuracy. 

A recent approach towards differentially private deep neural networks was explored by Phan et al. \cite{Phan0WD16}. This work proposed deep private auto-encoders (dPAs), in which differential privacy is enforced by perturbing the cross-entropy errors in auto-encoders \cite{Bengio2009}. 
Their algorithm was designed particularly for auto-encoders, in which specific objective functions are applied. A different method, named \textbf{CryptoNets}, was proposed in \cite{pmlr-v48-gilad-bachrach16} towards the application of neural networks to encrypted data. A data owner can send their encrypted data to a cloud service that hosts the network, and get encrypted predictions in return. This method is different from our context, since it does not aim at releasing learning models under privacy protections.

\begin{wrapfigure}{l}{0.2\textwidth}
\vspace{-5pt}
  \begin{center}
    \includegraphics[width=0.2\textwidth]{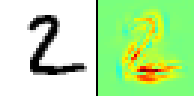} \vspace{-10pt}
    \caption{An instance of relevance of each input feature given to the classification output (MNIST dataset). Red neurons indicate stronger relevances, and green neurons indicate weaker relevances.}
    \label{LRP}
  \end{center}
\end{wrapfigure}

Another drawback of the existing techniques is that all parameters are treated the same in terms of the amount of noise injected. This may not be ideal in real scenarios, since different features and parameters normally have different impacts upon the model output. Fig. \ref{LRP} shows the relevance estimated by applying Layer-wise Relevance Propagation (LRP) \cite{bach-plos15} of input features to the prediction of the image's label in the MNIST dataset \cite{Lecun726791}. Dark red units have stronger impacts than green and yellow units. Injecting the same magnitude of noise into all parameters may affect the model utility. 

\textbf{Research Goal.} Therefore, there is an urgent demand for the development of a privacy preserving mechanism, such that: \textbf{(1)} It is totally independent of the number of training epochs in consuming privacy budget; \textbf{(2)} It has the ability to adaptively inject noise into features based on the contribution of each to the model output; and \textbf{(3)} It can be applied in a variety of deep neural networks. Mechanisms with such characteristics will significantly enhance the operation of privacy preservation in deep learning. 

\textbf{Our Contribution.} Motivated by this, we develop a novel mechanism, called \textbf{Ad}aptive \textbf{L}aplace \textbf{M}echanism (\textbf{AdLM}), to preserve differential privacy in deep learning. Our idea is to intentionally add \textit{``more noise"} into features which are \textit{``less relevant"} to the model output, and vice-versa. To achieve that, we inject Laplace noise into the computation of Layer-wise Relevance Propagation (LRP) \cite{bach-plos15} to estimate a differentially private relevance of each input feature to the model output. Given the perturbed features, we figure out a novel way to distribute adaptive noise into affine transformations and loss functions used in deep neural networks as a preprocessing step, so that preserving differential privacy is feasible. As a result, we expect to improve the utility of deep neural networks under $\epsilon$-differential privacy. It is worth noting that our mechanism does not access the original data again in the training phase. Theoretical analysis derives the sensitivities and error bounds of our mechanism, and shows that they are totally independent of the number of epochs. 

Different from \cite{Abadi,ShokriVitaly2015}, in our mechanism, the injected noise and the privacy budget consumption do not accumulate in each training step. Consequently, the privacy budget consumption in our mechanism is totally independent of the number of training epochs. In addition, different from \cite{Phan0WD16}, our mechanism can be applied in a variety of deep learning networks with different activation functions. Convolution neural networks (\textbf{CNNs}) \cite{Lecun726791} are used as an example to validate the effectiveness of our mechanism. Rigorous experiments conducted on MNIST and CIFAR-10 datasets \cite{krizhevsky2009learning} show that our mechanism is effective and outperforms existing solutions. 


\section{Preliminaries and Related Works}
In this section, we revisit differential privacy, existing techniques in preserving differential privacy in deep learning, and the Layer-wise Relevance Propagation (LRP) algorithm \cite{bach-plos15}. 

Let $D$ be a database that contains $n$ tuples $\mathbf{x}_1, \mathbf{x}_2, \ldots, \mathbf{x}_n$ and $d$+$1$ attributes $X_1, X_2, \ldots, X_d, Y$, and for each tuple $\mathbf{x}_i = (x_{i1}, x_{i2}, \ldots, x_{id}, y_i)$. We assume, without loss of generality, $\sqrt{\sum^d_{j=1} x^2_{ij}} \leq 1$ where $x_{ij} \geq 0$. This assumption can be easily enforced by changing each $x_{ij}$ to $\frac{x_{ij}-\alpha_j}{(\beta_j-\alpha_j)\cdot\sqrt{d}}$, where $\alpha_j$ and $\beta_j$ denote the minimum and maximum values in the domain of $X_j$. 

To be general, let us consider a classification task with $M$ possible categorical outcomes, i.e., the data label $y_i$ given $\mathbf{x}_i \in L$ is assigned to only one of the $M$ categories. Each $y_i$ can be considered as a vector of $M$ categories $y_i = \{y_{i1}, \ldots, y_{iM}\}$. If the $l$-th category is the class of $\mathbf{x}_i$, then $y_{il} = 1$, otherwise $y_{il} = 0$. 
Our objective is to construct a differentially private deep neural network from $D$ that (i) takes $\mathbf{x}_i = (x_{i1}, x_{i2}, \ldots, x_{id})$ as input and (ii) outputs a prediction of $y_i$ that is as accurate as possible. To evaluate whether model parameters $\theta$ lead to an accurate model, a cost function $\mathcal{F}_D(\theta)$ is used to measure the difference between the original and predicted values of $y_i$.

\subsection{$\epsilon$-Differential Privacy}
As the released model parameter $\theta$ may disclose sensitive information of $D$, to protect the privacy, we require that the model training should be performed with an algorithm that satisfies $\epsilon$\textit{-differential privacy}. The definition of differential privacy is as follows: 

\begin{mydef}{$\epsilon$-Differential Privacy \cite{dwork2006calibrating}.} A randomized algorithm $A$ fulfills $\epsilon$-differential privacy, if for any two databases $D$ and $D'$ differing at most one tuple, and for all $O \subseteq Range(A)$, we have: 
\begin{equation} \small
Pr[A(D) = O] \leq e^\epsilon Pr[A(D') = O]  
\end{equation}
where the privacy budget $\epsilon$ controls the amount by which the distributions induced by $D$ and $D'$ may differ. A smaller $\epsilon$ enforces a stronger privacy guarantee of $A$.
\label{Different Privacy} 
\end{mydef}

A general method for preserving $\epsilon$-differential privacy of any function $\mathcal{F}$ (on $D$) is the \textit{Laplace mechanism} \cite{dwork2006calibrating}, where the output of $\mathcal{F}$ is a vector of real numbers. In fact, the mechanism exploits the global sensitivity of $\mathcal{F}$ over any two neighboring data sets (differing at most one record), which is denoted as $GS_{\mathcal{F}}(D)$. Given $GS_{\mathcal{F}}(D)$, the Laplace mechanism ensures $\epsilon$-differential privacy by injecting noise $\eta$ into each value in the output of $\mathcal{F}(D)$: 
$pdf(\eta) = \frac{\epsilon}{2GS_{\mathcal{F}}(D)}exp(-|\eta|\cdot \frac{\epsilon}{GS_{\mathcal{F}}(D)})$, 
where $\eta$ is drawn i.i.d. from Laplace distribution with zero mean and scale $GS_{\mathcal{F}}(D)/\epsilon$.

Research in differential privacy has been significantly studied, from both the theoretical perspective, e.g.,
\cite{nipsChaudhuriM08,DBLP:conf/sigmod/KiferM11}, and the application perspective, e.g., data collection \cite{RAPPOR}, spatio-temporal correlations \cite{2016Changchang,DBLPCaoYX016}, data streams \cite{Chan:2012}, stochastic gradient descents \cite{SongCS13}, recommendation \cite{mcsherry2009differentially}, regression \cite{nipsChaudhuriM08}, online learning \cite{JainKT12}, publishing contingency tables \cite{xiao2010differential}, and spectral graph analysis \cite{DBLP:conf/pakdd/WangWW13}.

\subsection{Differential Privacy in Deep Learning}
\begin{wrapfigure}{l}{0.16\textwidth} 
\vspace{-10pt}
  \begin{center}
    \includegraphics[width=0.16\textwidth]{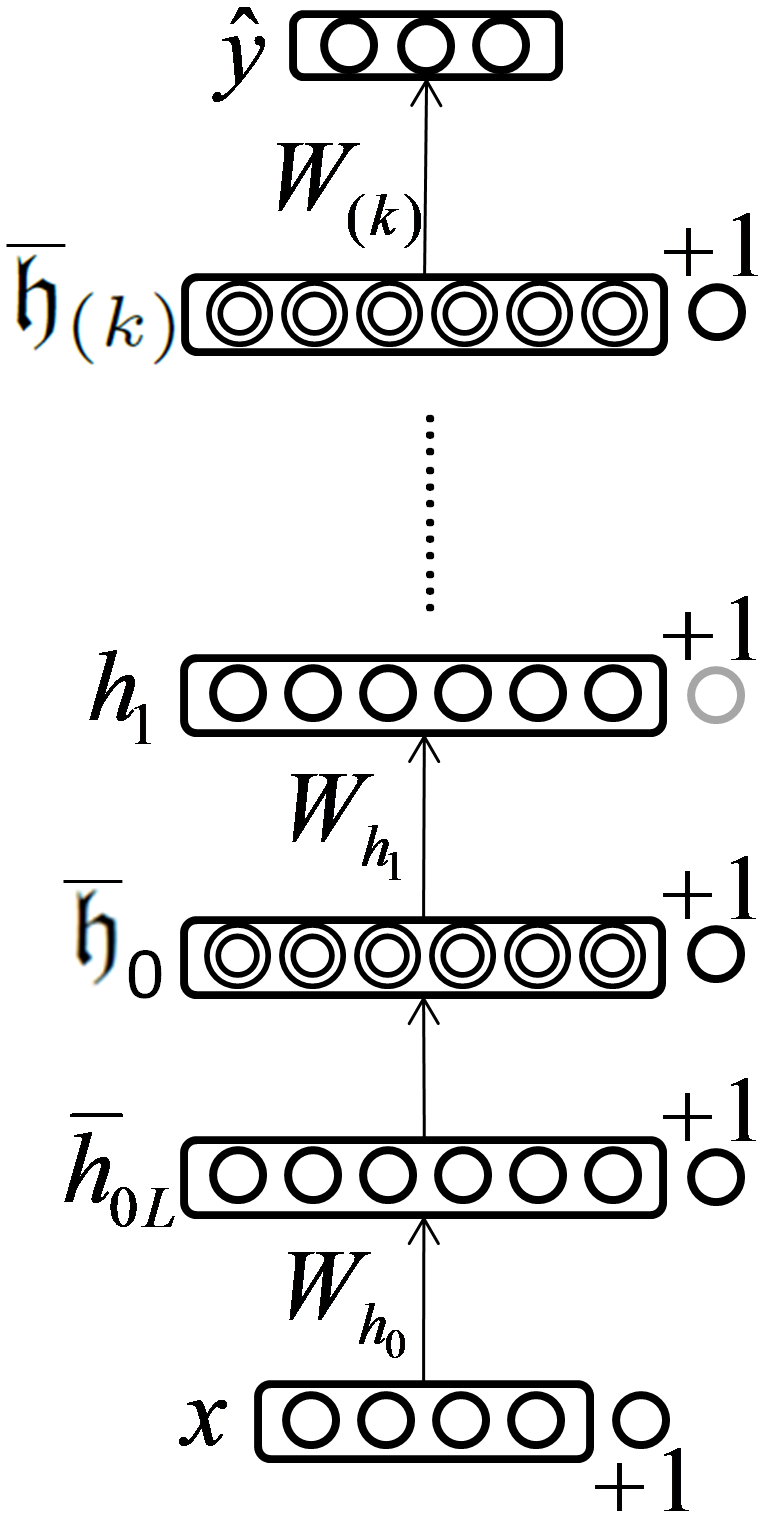} \vspace{-15pt}
    \caption{An instance of differentially private neural networks.} \vspace{-5pt}
    \label{DiPU}
  \end{center}
\end{wrapfigure}

Deep neural networks define parameterized functions from inputs $\mathbf{x}_i \in D$ to outputs, i.e, a prediction of $y_i$, as compositions of many layers of hidden neurons and nonlinear functions. For instance, Fig. \ref{DiPU} illustrates a multilayer neural network, in which there are $k$ hidden layers $H = \{\mathbf{h}_1, \ldots, \mathbf{h}_k\}$. Rectified linear units (ReLUs) and sigmoids are widely used examples of activation functions. By adjusting parameters of these neurons, such parameterized functions can be trained with the goal of fitting a finite set of input-output data instances. We specify a loss function $\mathcal{F}_D(\theta)$ that represents the penalty for mismatching between the predicted and original values of $y_i$. $\mathcal{F}_D(\theta)$ on parameters $\theta$ is the average of the loss over the training examples $\{\mathbf{x}_1,$...$, \mathbf{x}_{n}\}$. Stochastic gradient descent (SGD) algorithm is used to minimize the cross-entropy error \cite{Bengio2009}, given the model outputs and true data labels.

In the work of Abadi et al. \cite{Abadi}, to preserve differential privacy, normal (Gaussian) distribution noise is added into the gradients $\tilde{g}$ of parameters $W$ as follows. At each training step $t$, the algorithm first takes a random sample $L_t$ with sampling probability $L / n$, where $L$ is a group size and $n$ is the number of tuples in $D$. For each tuple $\mathbf{x}_i \in L_t$, the gradient $\mathbf{g}_t(\mathbf{x}_i) = \nabla_{\theta_t}\mathcal{F}_{\mathbf{x}_i}(\theta_t)$ is computed. Then the gradients will be bounded by clipping each gradient in $l_2$ norm, i.e., the gradient vector $\mathbf{g}_t$ is replaced by $\mathbf{g}_t/\max(1, \lVert \mathbf{g}_t^2 \rVert/C)$ for a predefined threshold $C$. Normal distribution noise is added into gradients of parameters $\theta$ as
\begin{align} 
\tilde{g}_t &\leftarrow \frac{1}{L}\sum_{i} \Big(\frac{\mathbf{g}_t(\mathbf{x}_i)}{\max(1, \frac{\lVert \mathbf{g}_t(\mathbf{x}_i)^2 \rVert}{C})} + \mathcal{N}(0, \sigma^2 C^2 \mathbf{I}) \Big) 
\\ 
\theta_t &\leftarrow \theta_t - \xi_t \tilde{g}_t 
\end{align}
where $\xi_t$ is a learning rate at the training step $t$.

Finally, differentially private parameters $\theta$, denoted $\overline{\theta}$, learned by the algorithm are shared to the public and other parties. Overall, the algorithm introduces noise into \textit{``gradients"} of parameters at every training step. The magnitude of injected noise and the privacy budget $\epsilon$ are accumulated in proportion to the number of training epochs. 

\textbf{Compared with the work in \cite{Abadi}}, the goal is similar: learning differentially private parameters $\overline{\theta}$. However, we develop a novel mechanism in which the privacy budget consumption is independent of the number of training epochs. Our mechanism is different. We redistribute the noise so that \textit{``more noise"} will be added into features which are \textit{``less relevant"} to the model output, and vice-versa. Moreover, we inject noise into coefficients of affine transformations and loss functions, such that differentially private parameters can be learned. 

\subsection{Layer-wise Relevance Propagation}
Layer-wise Relevance Propagation (LRP) \cite{bach-plos15} is a well-accepted algorithm, which is applied to compute the relevance of each input feature $x_{ij}$ to the model outcome $\mathcal{F}_{\mathbf{x}_i}(\theta)$. Given the relevance, denoted $R_{m}^{(k)}(\mathbf{x}_i)$, of a certain neuron ${m}$ at the layer $k$, i.e., ${m} \in \mathbf{h}_{k}$, for the model outcome $\mathcal{F}_{\mathbf{x}_i}(\theta)$, LRP algorithm aims at obtaining a decomposition of such relevance in terms of messages sent to neurons of the previous layers, i.e., the layer $(k$-$1)$-th. These messages are called $R^{(k-1, k)}_{p \leftarrow m}(\mathbf{x}_i)$. The overall relevance of each neuron in the lower layer is determined by summing up the relevance coming from all upper-layer neurons:
\begin{equation}
R_{p}^{(k-1)}(\mathbf{x}_i) = \sum_{m \in \mathbf{h}_{k}} R_{p \leftarrow m}^{(k-1, k)}(\mathbf{x}_i)
\label{Relevance2} 
\end{equation}
where the relevance decomposition is based on the ratio of local and global affine transformations and is given by:
\begin{equation} 
R_{p \leftarrow m}^{(k-1, k)} (\mathbf{x}_i) = 
\left \{
  \begin{tabular}{c}
  $\frac{z_{pm}(\mathbf{x}_i)}{z_m(\mathbf{x}_i) + \mu} R_{m}^{(k)}(\mathbf{x}_i)$ \; $z_m(\mathbf{x}_i) \geq 0$ \\
  $\frac{z_{pm}(\mathbf{x}_i)}{z_m(\mathbf{x}_i) - \mu} R_{m}^{(k)}(\mathbf{x}_i)$ \; $z_m(\mathbf{x}_i) < 0$ 
  \end{tabular}
\right. 
\label{Relevance1}
\end{equation}
with: $z_m(\mathbf{x}_i)$ is the affine transformation of neuron $m \in \mathbf{h}_{k}$:
\begin{align}
z_{pm}(\mathbf{x}_i) &= p_{\mathbf{x}_i} \times W_{pm} 
\\
z_m(\mathbf{x}_i) &= \sum_{p \in \mathbf{h}_k} z_{pm}(\mathbf{x}_i) + b_m
\end{align}
s.t. $p_{\mathbf{x}_i}$ is the value of neuron $p$ given $\mathbf{x}_i$, $W_{pm}$ is a weight connecting the neuron $p$ to neuron $m$, and $b_m$ is a bias term. A predefined stabilizer $\mu \geq 0$ is introduced to overcome unboundedness. 

In Eq. \ref{Relevance1}, in order to back propagate the relevance, we need to compute the relevance $R_{m}^{(k)}(\mathbf{x}_i)$ at the last hidden layer, i.e., the $k$-th layer, from the output layer. Given the output variable $o$, $R_{m}^{(k)}(\mathbf{x}_i)$ is computed as follows: 
\begin{equation}
R_{m}^{(k)} (\mathbf{x}_i) = 
\left \{
  \begin{tabular}{c}
  $\frac{z_{mo}(\mathbf{x}_i)}{z_o(\mathbf{x}_i) + \mu} \mathcal{F}_{\mathbf{x}_i}(\theta)$ \; $z_o(\mathbf{x}_i) \geq 0$ \\
  $\frac{z_{mo}(\mathbf{x}_i)}{z_o(\mathbf{x}_i) - \mu} \mathcal{F}_{\mathbf{x}_i}(\theta)$ \; $z_o(\mathbf{x}_i) < 0$ 
  \end{tabular}
\right. 
\label{Relevance3}
\end{equation}

Given $k$ hidden layers $\{\mathbf{h}_1, \ldots, \mathbf{h}_k\}$, by using Eqs. \ref{Relevance2}, \ref{Relevance1}, and \ref{Relevance3}, we can compute the relevance of every hidden neuron and input feature. As in \cite{bach-plos15}, the following equation holds:
\begin{multline}
\mathcal{F}_{\mathbf{x}_i}(\theta) = \sum_{m \in \mathbf{h}_{k}} R_{m}^{(k)}(\mathbf{x}_i) 
= \ldots 
= \sum_{x_{ij} \in \mathbf{x}_{i}} R_{x_{ij}}(\mathbf{x}_i)
\end{multline}
where $R_{x_{ij}}(\mathbf{x}_i)$ is the relevance of the feature $x_{ij}$ given the model outcome $\mathcal{F}_{\mathbf{x}_i}(\theta)$. To ensure that the relevance $R_{x_{ij}}(\mathbf{x}_{i}) \in [-1, 1]$, each $R_{x_{ij}}(\mathbf{x}_{i})$ is normalized to $\frac{R_{x_{ij}}(\mathbf{x}_{i}) - \chi}{(\varphi - \chi)}$, where $\varphi$ and $\chi$ denote the maximum and minimum values in the domain of $\big\{R_{x_{i1}}(\mathbf{x}_{i}), \ldots, R_{x_{id}}(\mathbf{x}_{i})\big\}$.

\section{Adaptive Laplace Mechanism (AdLM)}
In this section, we formally present our mechanism. Given a loss function $\mathcal{F}(\theta)$ with model parameters $\theta$, the network is trained by optimizing the loss function $\mathcal{F}(\theta)$ on $D$ by applying SGD algorithm on $T$ random training batches consequently. At each training step, a single training batch $L$ is used. A batch $L$ is a random set of training samples in $D$ with a predefined batch size $|L|$. 

The pseudo-codes of Algorithm \ref{AdLM} outline five basic steps in our mechanism to learn differentially private parameters of the model. The five basic steps are as follows:

$\bullet$ \textbf{Step 1 (Lines 1-7).} In the first step, we obtain the average relevances of all the $j$-th input features, denoted as $R_j(D)$, by applying the LRP algorithm on a well-trained deep neural network on the database $D$. $R_j(D)$ is computed as follows: \vspace{-2pt}
\begin{equation}
R_j(D) = \frac{1}{|D|} \sum_{\mathbf{x}_{i} \in D} R_{x_{ij}}(\mathbf{x}_i) \vspace{-2pt}
\label{R_j(D)}
\end{equation}
Then, we derive differentially private relevances, denoted as $\overline{R}_j$, by injecting Laplace noise into $R_j$ for all the $j$-th input features. The total privacy budget in this step is $\epsilon_1$. 

$\bullet$ \textbf{Step 2 (Lines 8-14).} In the second step, we derive a differentially private affine transformation layer, denoted $\mathbf{h}_{0}$. Every hidden neuron $h_{0j} \in \mathbf{h}_0$ will be perturbed by injecting adaptive Laplace noise into its affine transformation to preserve differential privacy given a batch $L$. Based on $\overline{R}_j$, \textit{``more noise"} is injected into features which are \textit{``less relevant"} to the model output, and vice-versa. The total privacy budget used in this step is $\epsilon_2$. The perturbed affine transformation layer is denoted as $\overline{\mathbf{h}}_{0L}$ (Fig. \ref{DiPU}). 

$\bullet$ \textbf{Step 3 (Line 15).} In the third step, we stack hidden layers $\{\mathbf{h}_1, \ldots, \mathbf{h}_k\}$ on top of the differentially private hidden layer $\overline{\mathbf{h}}_{0L}$ to construct the deep private neural network (Fig. \ref{DiPU}). The computations of $\mathbf{h}_1, \ldots, \mathbf{h}_k$ are done based on the differentially private layer $\overline{\mathbf{h}}_{0L}$ without accessing any information from the original data. Therefore, the computations do not disclose any information. Before each stacking operation, a normalization layer, denoted $\overline{\mathfrak{h}}$, is applied to bound non-linear activation functions, such as ReLUs (Fig. \ref{DiPU}). 

$\bullet$ \textbf{Step 4 (Lines 16-19).} After constructing a private structure of hidden layers $\{\overline{\mathbf{h}}_{0L}, \mathbf{h}_1, \ldots, \mathbf{h}_k\}$, we need to protect the labels $y_i$ at the output layer. To achieve this, we derive a polynomial approximation of the loss function $\mathcal{F}$. Then, we perturb the loss function $\mathcal{F}$ by injecting Laplace noise with a privacy budget $\epsilon_3$ into its coefficients to preserve differential privacy on each training batch $L$, denoted $\overline{\mathcal{F}}_L(\theta)$. 

$\bullet$ \textbf{Step 5 (Lines 20-30).} Finally, the parameter $\theta_T$ is derived by minimizing the loss function $\overline{\mathcal{F}}_L(\theta)$ on $T$ training steps sequentially. In each step $t$, stochastic gradient descent (SGD) algorithm is used to update parameters $\theta_t$ given a random batch $L$ of training samples in $D$. This essentially is an optimization process, without using any additional information from the original data. 

In our mechanism, differential privacy is preserved, since it is enforced at every computation task that needs to access the original data $D$. Laplace noise is injected into our model only once, as a preprocessing step to preserve differential privacy in the computation of the relevance $\overline{R}_j(D)$, the first layer $\overline{\mathbf{h}}_{0L}$, and the loss function $\overline{\mathcal{F}}_L(\theta)$. Thereafter, the training phase will not access the original data again. The privacy budget consumption does not accumulate in each training step. As such, it is independent of the number of training epochs.

\subsection{Private Relevance}
In this section, we preserve differential privacy in the computation of the relevance of each $j$-th input feature on database $D$ by injecting Laplace noise into $R_j(D)$. We set $\Delta_{\mathbf{R}} = \frac{2d}{|D|}$ based on the maximum values of all the relevances $R_j(D)$ (line 4, Alg. \ref{AdLM}). In lines 5-6, the relevance of each $j$-th input feature $R_j(D)$ is perturbed by adding Laplace noise $Lap(\frac{\Delta_{\mathbf{R}}}{\epsilon_1})$. The perturbed relevance is denoted as $\overline{R}_j$. In line 7, we obtain the set of all perturbed relevances $\overline{\mathbf{R}}(D)$:\vspace{-2pt}
\begin{align}
&\overline{\mathbf{R}}(D) = \big\{\overline{R}_j\big\}_{j \in [1, d]}  
\\
&\textit{where } \overline{R}_j = \frac{1}{|D|} \sum_{\mathbf{x}_{i} \in D} R_{x_{ij}}(\mathbf{x}_i) + Lap(\frac{\Delta_{\mathbf{R}}}{\epsilon_1}) \vspace{-2pt}
\label{DPRelevance}
\end{align}

The computation of $\overline{\mathbf{R}}(D)$ is $\epsilon_1$-differential private. The correctness is based on the following lemmas. 
\begin{lemma}
Let $D$ and $D'$ be any two neighboring databases. Given $\mathbf{R}(D)$ and $\mathbf{R}(D')$ be the relevance of all input features on $D$ and $D'$, respectively, and denote their representations as \vspace{-10pt}
\begin{align}
\mathbf{R}(D) &= \big\{R_j(D)\big\}_{j \in [1, d]} \textit{\ s.t. } R_j(D) = \frac{1}{|D|} \sum_{\mathbf{x}_{i} \in D} R_{x_{ij}}(\mathbf{x}_i) \nonumber 
\\ 
\mathbf{R}(D') &= \big\{R_j(D')\big\}_{j \in [1, d]} \textit{\ s.t. } R_j(D') = \frac{1}{|D'|} \sum_{\mathbf{x}'_{i} \in D'} R_{x'_{ij}}(\mathbf{x}'_i) \nonumber
\end{align}
Then, we have the following inequality: \vspace{-2pt}
\begin{multline} \small
\frac{1}{|D|} \sum_{j = 1}^d \Big\lVert \sum_{\mathbf{x}_i \in D} R_{x_{ij}}(\mathbf{x}_i) - \sum_{\mathbf{x}'_i \in D'} R_{x'_{ij}}(\mathbf{x}'_i) \Big\rVert_1 
\leq \frac{2d}{|D|}
\label{GlobalSensitivity}
\end{multline}
where $d$ is the number of features in each tuple $\mathbf{x}_i \in D$.
\label{Lemma1}
\end{lemma}

\textit{Proof 1: } The proof of Lemma \ref{Lemma1} is in \textbf{Appendix A}. 

\begin{lemma} Algorithm \ref{AdLM} preserves $\epsilon_1$-differential privacy in the computation of $\overline{\mathbf{R}}(D)$. 
\label{Lemma2}
\end{lemma}

\textit{Proof 2: } The proof of Lemma \ref{Lemma2} is in \textbf{Appendix B}$^1$.

\begin{wrapfigure}{l}{0.17\textwidth} 
\vspace{-5pt}
  \begin{center}
    \includegraphics[width=0.17\textwidth]{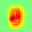} \vspace{-15pt}
    \caption{The average differentially private relevance of each input feature given MNIST dataset.}
    \label{LRP_D}
  \end{center}
\vspace{-5pt}
\end{wrapfigure}

Lemma \ref{Lemma2} shows that the computation of the relevances $\overline{\mathbf{R}}(D)$ is differentially private. Fig. \ref{LRP_D} illustrates the differentially private relevance $\overline{R}_j$ (Eq. \ref{DPRelevance}) of each $j$-th coefficient given the database $D$. It is worth noting that the relevance distribution is not identical. In the next section, $\overline{\mathbf{R}}(D)$ is used to redistribute the noise injected into the affine transformation layer $\mathbf{h}_0$ in our deep neural network.

\subsection{Private Affine Transformation Layer with Adaptive Noise}
In general, before applying activation functions such as ReLU and sigmoid, the affine transformation of a hidden neuron $h \in \mathbf{h}_0$ can be presented as: \vspace{-2pt}
\begin{equation}
h_{\mathbf{x}_i}(W) = b + \mathbf{x}_i W^T \vspace{-2pt}
\label{LReLU}
\end{equation}
where $b$ is a static bias, and $W$ is the parameter of $h$. 
Given a training batch $L$, $h$ can be rewritten as: \vspace{-2pt}
\begin{equation}
h_{L}(W) = \sum_{\mathbf{x}_i \in L} (b + \mathbf{x}_i W^T) \vspace{-2pt}
\end{equation}

Given the above representation of each neuron $h_{L}(W_{h})$, 
we preserve differential privacy in the computation of $\mathbf{h}_0$ on $L$ by injecting Laplace noise into inputs $b$ and $\mathbf{x}_i$ of every neuron $h_{L}(W) \in \mathbf{h}_0$. Intuitively, we can apply an identical noise distribution $\frac{1}{|L|}Lap(\frac{\Delta_{\mathbf{h}_0}}{\epsilon_2})$ to all input features, where $\Delta_{\mathbf{h}_0} = 2 \sum_{h \in \mathbf{h}_0} d$ (line 9, Alg. \ref{AdLM}). This approach works well when every input feature has an identical contribution to the model outcome. (Please refer to the \textbf{Appendix C}$^1$.)

In practice, this assumption usually is violated. For instance, Fig. \ref{LRP} illustrates the relevance, estimated by the LRP algorithm \cite{bach-plos15}, of each input feature given different handwritten digits. It is clear that the relevances are not identical. The differentially private relevances are not identical as well (Fig. \ref{LRP_D}). Therefore, injecting the same magnitude of noise into all input features may affect the utility of differentially private neural networks. 

To address this problem, we propose an Adaptive Laplace Mechanism (\textbf{AdLM}), to adaptively redistribute the injected noise to improve the performance. Given hidden units $h_{\mathbf{x}_i}(W)$ in Eq. \ref{LReLU}, our key idea is to intentionally add \textit{more noise} into input features which are \textit{less relevant} to the model output $Y$, and vice-versa. As a result, we expect to improve the utility of the model under differential privacy. In fact, we introduce a privacy budget ratio $\beta_j$ and the privacy budget $\epsilon_j$ for each $j$-th input feature as follows: \vspace{-2.5pt}
\begin{equation}
\beta_j = \frac{d \times |\overline{R}_j|}{\sum_{j = 1}^d |\overline{R}_j|} \textit{\ \ \ s.t.\ \ \ } \epsilon_j = \beta_j \times \epsilon_2 \vspace{-2.5pt}
\label{BudgetRatio}
\end{equation}

We set $\Delta_{\mathbf{h}_0} = 2 \sum_{h \in \mathbf{h}_0}d$ \ based on the maximum values of all the input features ${x_{ij}}$ (line 9, Alg. \ref{AdLM}).
In line 11, $\beta_j$ can be considered as the fraction of the contribution to $\Delta_{\mathbf{h}_0}$ from the $j$-th input feature to the hidden neuron $h \in \mathbf{h}_0$. In lines 12-13, each input feature $x_{ij}$ of every hidden neuron $h$ in the first affine transformation layer $\mathbf{h}_0$ is perturbed by adding adaptive Laplace noise $\frac{1}{|L|}Lap(\Delta_{\mathbf{h}_0}/\epsilon_j)$. The perturbed input features are denoted as $\overline{\mathbf{x}}_i$. In lines 20-21, given a random training batch $L$, we construct the differentially private affine transformation layer $\overline{\mathbf{h}}_{0L}$, which consists of perturbed hidden neurons $\overline{h}_L(W)$: \vspace{-2.5pt}
\begin{equation}
\overline{\mathbf{h}}_{0L}(W_{0}) = \big\{\overline{h}_L(W)\big\}_{h \in \mathbf{h}_0} \textit{ s.t. } \overline{h}_L(W) = \sum_{\mathbf{x}_i \in L} \Big(\overline{\mathbf{x}}_i W^T + \overline{b}\Big) \nonumber \vspace{-2.5pt}
\end{equation}
where $\overline{b} = b + \frac{1}{|L|}Lap(\frac{\Delta_{\mathbf{h}_0}}{\epsilon_2})$ is the perturbed bias (line 14). The following lemma shows that Alg. \ref{AdLM} preserves $\epsilon_2$-differential privacy in the computation of $\overline{\mathbf{h}}_{0L}$. 
\begin{lemma}
Let $L$ and $L'$ be any two neighboring batches. Given parameter $W_0$, let $\mathbf{h}_{0L}$ and $\mathbf{h}_{0L'}$ be the first affine transformation layers on $L$ and $L'$, respectively, and denote their representations as follows: \vspace{-5pt}
\begin{align}
\mathbf{h}_{0L}(W_0) &= \{ h_{L}(W)\}_{h \in \mathbf{h}_0} \textit{ s.t. } h_{L}(W) = \sum_{\mathbf{x}_i \in L} (b + \mathbf{x}_{i} W^T) \nonumber 
\\ 
\mathbf{h}_{0L'}(W_{0}) &= \{ h_{L'}(W)\}_{h \in \mathbf{h}_0} \textit{ s.t. } h_{L'}(W) = \sum_{\mathbf{x}'_i \in L'} (b + \mathbf{x}'_{i} W^T) \nonumber
\end{align}
Then, we have the following inequality: \vspace{-5pt}
\begin{equation}
\Delta_{\mathbf{h}_0} = \sum_{h \in \mathbf{h}_0} \sum_{j = 1}^d \Big\lVert \sum_{\mathbf{x}_i \in L} {x}_{ij} - \sum_{\mathbf{x}'_i \in L'} {x}'_{ij} \Big\rVert_1 
\leq 2 \sum_{h \in \mathbf{h}_0} d
\label{GlobalSensitivity2}
\end{equation}
where $d$ is the number of features in each tuple $\mathbf{x}_i \in D$. 
\label{Lemma3}
\end{lemma}

\begin{algorithm}[t]
\caption{\textbf{Adaptive Laplace Mechanism} (Database $D$, hidden layers $H$, loss function $\mathcal{F}(\theta)$, and privacy budgets $\epsilon_1$, $\epsilon_2$, and $\epsilon_3$, the number of batches $T$, the batch size $|L|$)}
\label{AdLM}
\begin{algorithmic}[1]
\small{
\STATE \textbf{Compute the average relevance by applying the LRP Alg.}
\STATE $\forall j \in [1, d]: R_j(D) = \frac{1}{|D|} \sum_{\mathbf{x}_{i} \in D} R_{x_{ij}}(\mathbf{x}_i)$ \#Eq.\ref{R_j(D)}\#
\STATE \textbf{Inject Laplace noise into the average relevance of each $j$-th input feature}
\STATE $\Delta_{\mathbf{R}} = 2d/|D|$ \#Lemma \ref{Lemma1}\#
\FOR {$j \in [1,d]$}
       \STATE $\overline{R}_j \leftarrow \frac{1}{|D|} \sum_{\mathbf{x}_{i} \in D} R_{x_{ij}}(\mathbf{x}_i) + Lap(\frac{\Delta_{\mathbf{R}}}{\epsilon_1})$
\ENDFOR
\STATE $\overline{\mathbf{R}}(D) = \{\overline{R}_j\}_{j \in [1, d]}$
\STATE \textbf{Inject Laplace noise into coefficients of the differentially private layer $\mathbf{h}_0$}
\STATE $\Delta_{\mathbf{h}_0} = 2 \sum_{h \in \mathbf{h}_0} d$ \#Lemma \ref{Lemma3}\#
\FOR {$j \in [1, d]$}
         \STATE $\epsilon_j \leftarrow \beta_j \times \epsilon_2$ \#Eq. \ref{BudgetRatio}\#
\ENDFOR
         \FOR {$\mathbf{x}_i \in D, j \in [1, d]$}
                 \STATE $\overline{x}_{ij} \leftarrow x_{ij} + \frac{1}{|L|}Lap(\frac{\Delta_{\mathbf{h}_0}}{\epsilon_j})$ \#\textit{perturb input feature $x_{ij}$}\#
         \ENDFOR
\STATE $\overline{b} \leftarrow b + \frac{1}{|L|}Lap(\frac{\Delta_{\mathbf{h}_0}}{\epsilon_2})$ \#\textit{perturb bias $b$}\#
\STATE \textbf{Construct hidden layers $\{\mathbf{h}_1, \ldots, \mathbf{h}_k\}$ and normalization layers $\{\overline{\mathfrak{h}}_1, \ldots, \overline{\mathfrak{h}}_{(k)}\}$}
\STATE \textbf{Inject Laplace noise into coefficients of the approximated loss function $\widehat{\mathcal{F}}$}
\STATE $\Delta_{\mathcal{F}} = M(|\overline{\mathfrak{h}}_{(k)}| + \frac{1}{4}|\overline{\mathfrak{h}}_{(k)}|^2)$ \#Lemma \ref{lemma5}\#
\FOR {$\mathbf{x}_i \in D, R \in [0,2], l \in [1,M]$}
         \STATE $\overline{\phi}^{(R)}_{l \mathbf{x}_i} \leftarrow \phi^{(R)}_{l \mathbf{x}_i} + \frac{1}{|L|}Lap(\frac{\Delta_{\mathcal{F}}}{\epsilon_3})$ \#\textit{perturb coefficients of $\widehat{\mathcal{F}}$}
\ENDFOR
\STATE \textbf{Initialize} $\theta_0$ randomly
\FOR {$t \in [T]$}
         \STATE \textbf{Take a random training batch $L$}
	  \STATE \textbf{Construct differentially private affine transformation layer}
         \STATE $\overline{\mathbf{h}}_{0L}(W_0) \leftarrow \{\overline{h}_L(W)\}_{h \in \mathbf{h}_0}$ 
         \STATE \textit{s.t. } $\overline{h}_L(W) = \sum_{\mathbf{x}_i \in L} \big(\overline{\mathbf{x}}_i W^T + \overline{b})\big]\big)$
         \STATE \textbf{Construct differentially private loss function}
         \STATE $\overline{\mathcal{F}}_L(\theta_t) = \sum_{l = 1}^M \sum_{\mathbf{x}_i \in L} \sum_{R = 0}^{2} \big(\overline{\phi}^{(R)}_{l \mathbf{x}_i} W_{l(k)}^T\big)^R$
         \STATE \textbf{Compute gradient descents}
         \STATE $\theta_{t+1} \leftarrow \theta_t - \eta_t \frac{1}{|L|} \triangledown_{\theta_t}\overline{\mathcal{F}}_L(\theta_t)$ \#$\eta_t$ is a learning rate\#
\ENDFOR
\STATE \textbf{Return} $\theta_T$  \#\textit{$(\epsilon_1 + \epsilon_2 + \epsilon_3)$-differentially private}\#
}
\end{algorithmic}
\end{algorithm}


\textit{Proof 3: }
Assume that $L$ and $L'$ differ in the last tuple. Let $\mathbf{x}_n$ ($\mathbf{x}'_n$) be the last tuple in $L$ ($L'$). We have that \vspace{-5pt}
\begin{align}
&\Delta_{\mathbf{h}_0} = \sum_{h \in \mathbf{h}_0} \sum_{j = 1}^d \Big\lVert \sum_{\mathbf{x}_i \in L} x_{ij} - \sum_{\mathbf{x}'_i \in L'} x'_{ij} \Big\rVert_1 \nonumber \\
&= \sum_{h \in \mathbf{h}_0} \sum_{j = 1}^d \lVert x_{nj} - x'_{nj} \rVert_1 \leq 2 \max_{\mathbf{x}_i \in L} \sum_{h \in \mathbf{h}_0} \sum_{j = 1}^d \lVert x_{ij} \rVert_1
\label{ProofLemma2}
\end{align}
Since $\forall \mathbf{x}_i, j: x_{ij} \in [0, 1]$, from Eq. \ref{ProofLemma2} we have that: $\Delta_{\mathbf{h}_0} \leq 2 \sum_{h \in \mathbf{h}_0} d$. Eq. \ref{GlobalSensitivity2} holds. \ \ \ \ \ \ \ \ \ \ \ \ \ \ \ \ \ \ \ \ \ \ \ \ \ \ \ \ \ \ \ \ \ \ \ \ \ \ \ \ $\blacksquare$

\begin{lemma} Algorithm \ref{AdLM} preserves $\epsilon_2$-differential privacy in the computation of $\overline{\mathbf{h}}_{0L}(W_0)$ (lines 24-25). 
\label{lemma4}
\end{lemma}

\textit{Proof 4: }
From lines 24-25 in the Alg. \ref{AdLM}, for each $h \in \overline{h}_{0L}$, $h$ can be re-written as:
\begin{multline}
\overline{h}_{L}(W) = \sum_{j = 1}^d \Big[\sum_{\mathbf{x}_i \in L} \big(x_{ij} + \frac{1}{|L|}Lap(\frac{\Delta_{\mathbf{h}_0}}{\epsilon_j})\big) W^T \Big] \\
+ \sum_{\mathbf{x}_i \in L}\big(b + \frac{1}{|L|}Lap(\frac{\Delta_{\mathbf{h}_0}}{\epsilon_2})\big)
\label{perturbedh3}
\end{multline}

Let us consider the static bias $b=1$ as the $0$-th input feature and its associated parameter $W_b$, i.e., $x_{i0} = b = 1$ and $W = W_b \cup W$, we have that
\begin{align}
& \overline{h}_L(W) = \sum_{j =0}^d \Big[\sum_{\mathbf{x}_i \in L} \big(x_{ij} + \frac{1}{|L|}Lap(\frac{\Delta_{\mathbf{h}_0}}{\epsilon_j})\big) W^T \Big] \label{perturbedh4}
\\
& = \sum_{j = 0}^d \Big[\sum_{\mathbf{x}_i \in L} x_{ij} + Lap(\frac{\Delta_{\mathbf{h}_0}}{\epsilon_j})\Big]W^T = \sum_{j = 0}^d \overline{\phi}^h_j W^T
\label{perturbedh5}
\end{align}
where $\overline{\phi}^h_j = \big[\sum_{\mathbf{x}_i \in L} x_{ij} + Lap(\frac{\Delta_{\mathbf{h}_0}}{\epsilon_j})\big]$.

We can see that $\overline{\phi}^h_j$ is the perturbation of the input feature $x_{ij}$ associated with the $j$-th parameter $W_j \in W$ of the hidden neuron $h$ on $L$. Since all the hidden neurons $h$ in ${\mathbf{h}_0}$ are perturbed, we have that: 
\begin{equation}
Pr\big(\overline{\mathbf{h}}_{0L}(W_0)\big) = \prod_{h \in \mathbf{h}_0} \prod_{j = 0}^d exp\big(\frac{\epsilon_j \lVert \sum_{\mathbf{x}_i \in L}  x_{ij} -  \overline{\phi}^{h}_j\rVert}{\Delta_{\mathbf{h}_0}}\big) \nonumber 
\end{equation}

$\Delta_{\mathbf{h}_0}$ is set to $2 \sum_{h \in \mathbf{h}_0} d$ (line 9 in Alg. \ref{AdLM}). $\overline{\mathbf{h}}_{0L}(W_0)$ is the output (lines 24-25 in Alg. \ref{AdLM}). We have that
\begin{align}
&\frac{Pr\big(\overline{\mathbf{h}}_{0L}(W_0)\big)}{Pr\big(\overline{\mathbf{h}}_{0L'}(W_0)\big)} = \frac{\prod_{h \in \mathbf{h}_0} \prod_{j = 0}^d \exp\big(\frac{\epsilon_j \lVert \sum_{\mathbf{x}_i \in L}  x_{ij} -  \overline{\phi}^{h}_j\rVert_1}{\Delta_{\mathbf{h}_0}}\big)}{\prod_{h \in \mathbf{h}_0} \prod_{j = 0}^d \exp\big(\frac{\epsilon_j \lVert \sum_{\mathbf{x}'_i \in L'} {x'}_{ij} - \overline{\phi}^{h}_j\rVert_1}{\Delta_{\mathbf{h}_0}}\big)} \nonumber
\\
&\leq \prod_{h \in \mathbf{h}_0} \prod_{j = 0}^d \exp(\frac{\epsilon_j}{\Delta_{\mathbf{h}_0}} \Big\lVert \sum_{\mathbf{x}_i \in L} x_{ij} -  \sum_{\mathbf{x}'_i \in L'} {x'}_{ij} \Big\rVert_1) \nonumber 
\\
&\leq \prod_{h \in \mathbf{h}_0} \prod_{j = 1}^d \exp(\frac{\epsilon_j}{\Delta_{\mathbf{h}_0}} 2\max_{\mathbf{x}_n \in L} \big\lVert x_{nj} \big\rVert _1) \leq \prod_{h \in \mathbf{h}_0} \prod_{j = 1}^d \exp(\frac{2\epsilon_j}{\Delta_{\mathbf{h}_0}}) \nonumber 
\\
&\leq \prod_{h \in \mathbf{h}_0} \prod_{j = 1}^d \exp(\epsilon_2 \frac{2 \frac{d \times |\overline{R}_j|}{\sum_{j = 1}^d |\overline{R}_j|}}{\Delta_{\mathbf{h}_0}}) \nonumber  
\\
&\leq \exp(\epsilon_2 \frac{2 \sum_{h \in \mathbf{h}_0} d \big[\sum_{j = 1}^d \frac{|\overline{R}_j|}{\sum_{j = 1}^d |\overline{R}_j|}\big]}{\Delta_{\mathbf{h}_0}}) = \exp(\epsilon_2) \nonumber 
\end{align}
Consequently, the computation of $\overline{\mathbf{h}}_{0L}(W_0)$ preserves $\epsilon_2$-differential privacy in Alg. \ref{AdLM}. \ \ \ \ \ \ \ \ \ \ \ \ \ \ \ \ \ \ \ \ \ \ \ \ \ \ \ \ \ \ \ \ \ \ \ $\blacksquare$

Lemma \ref{lemma4} shows that we can redistribute the noise in the computation of the first hidden layer $\overline{\mathbf{h}}_{0L}$ under differential privacy. In addition, given a batch $L$, without accessing additional information from the original data, none of the computations on top of $\overline{\mathbf{h}}_{0L}$ risk the privacy protection under differential privacy. These computation tasks include the application of activation functions, e.g., ReLU and sigmoid, on $\overline{\mathbf{h}}_{0L}$, the computation of hidden layers $\mathbf{h}_1, \ldots, \mathbf{h}_k$, local response normalizations, drop-out operations, polling layers, etc. (line 15, Alg. \ref{AdLM}). This result can be applied to both fully-connected layers and convolution layers. In this paper, we applied ReLU on top of $\overline{\mathbf{h}}_{0L}$ and other layers $\mathbf{h}_1, \ldots, \mathbf{h}_{k}$. Local response normalization layers are used after the application of ReLUs in each hidden layer to bound ReLU functions.

\textbf{Local Response Normalization.} The hidden units of the lower layer will be considered as the input of the next layer (Fig. \ref{DiPU}). To ensure that this input is bounded $\overline{h}_{\mathbf{x}_i} \in [0, 1]$, as in \cite{krizhevsky2012imagenet,Phan0WD16}, we add a local response normalization (LRN) layer on top of each hidden layer. Given a fully-connected layer, as in \cite{Phan0WD16}, given an input $\mathbf{x}_i$, each perturbed neuron $\overline{h}_{\mathbf{x}_i}(W)$ can be directly normalized as follows: 
$
\overline{\mathfrak{h}}_{\mathbf{x}_i} \leftarrow \Big(\overline{h}_{\mathbf{x}_i}(W) - \chi\Big)/(\varphi - \chi)
$, 
where $\varphi$ and $\chi$ denote the maximum and minimum values in the domain of $\{\overline{h}_{\mathbf{x}_i}\}_{i \in L}$.

Given a convolution layer with a perturbed neuron $\overline{h}^k_{ij}$ at location $(i, j)$ in the $k$-th feature map, based on \cite{krizhevsky2012imagenet}, our local response normalization (LRN) is presented as follows: 
\begin{equation}
\overline{\mathfrak{h}}^{k}_{ij} \leftarrow \overline{h}^{k}_{ij}/\max\Big(\overline{h}^{k}_{ij}, \big(q + \alpha \sum_{m = \max (0, k-l/2)}^{\min (N-1, k + l/2)} (\overline{h}^{m}_{ij})^2 \big)^\beta \Big)
\label{LRN2}
\end{equation}
where the constants $q, l, \alpha,$ and $\beta$ are hyper-parameters, $N$ is the total number of feature maps. As in \cite{krizhevsky2012imagenet}, we used $q = 2, l = 5, \alpha = 10^{-4},$ and $\beta = 0.75$ in our experiments.

\subsection{Perturbation of the Loss Function $\mathcal{F}_L(\theta)$}
On top of our private deep neural network (Fig. \ref{DiPU}), we add an output layer with the loss function $\mathcal{F}_L(\theta)$ to predict $Y$. Since the loss function $\mathcal{F}_L(\theta)$ accesses the labels $y_i$ given $\mathbf{x}_i \in L$ from the data, we need to protect the labels $y_i$ at the output layer. First, we derive a polynomial approximation of the loss function based on Taylor Expansion \cite{tagkey1985}. Then, we inject Laplace noise into coefficients of the loss function $\mathcal{F}$ to preserve differential privacy on each training batch $L$.

The model output variables $\{\hat{y}_1, \ldots, \hat{y}_M\}$ are fully linked to the normalized highest hidden layer, denoted $\overline{\mathfrak{h}}_{(k)}$, by weighted connections $W_{(k)}$ (Fig. \ref{DiPU}). As common, the logistic function can be used as an activation function of the output variables. Given, $l$-th output variable $\hat{y}_l$ and $\mathbf{x}_i$, we have:
\begin{equation}
\hat{y}_{il} = \sigma \big(\overline{\mathfrak{h}}_{\mathbf{x}_i(k)} W_{l(k)}^T\big)
\end{equation} 
where $\overline{\mathfrak{h}}_{\mathbf{x}_i(k)}$ is the state of $\overline{\mathfrak{h}}_{(k)}$ derived from $\overline{\mathbf{h}}_{0\mathbf{x}_i}$ by navigating through the neural network.

\textit{Cross-entropy error} \cite{Bengio2009} can be used as a loss function. It has been widely used and applied in real-world applications \cite{Bengio2009}. Therefore, it is critical to preserve differential privacy under the use of the cross-entropy error function. Other loss functions, e.g., square errors, can be applied in the output layer, as well. In our context, the cross-entropy error function is given by: 
\begin{align}
\mathcal{F}_L(\theta) &= -\sum_{l = 1}^M \sum_{\mathbf{x}_i \in L} \Big( y_{il}\log \hat{y}_{il} + (1-y_{il})\log (1 - \hat{y}_{il}) \Big) \nonumber 
\\
&= - \sum_{l = 1}^M \sum_{\mathbf{x}_i \in L} \Big( y_{il}\log (1+e^{-\overline{\mathfrak{h}}_{\mathbf{x}_i(k)} W_{l(k)}^T}) \nonumber 
\\
&\textit{\ \ \ \ \ \ \ \ \ \ \ \ \ \ \ \ } + (1-y_{il})\log (1+e^{\overline{\mathfrak{h}}_{\mathbf{x}_i(k)} W_{l(k)}^T}) \Big)
\label{Softmax} 
\end{align}

Based on \cite{Phan0WD16} and Taylor Expansion \cite{tagkey1985}, we derive the polynomial approximation of $\mathcal{F}_L(\theta)$ as: 
\begin{align}
&\widehat{\mathcal{F}}_L(\theta) = \sum_{l = 1}^M \sum_{\mathbf{x}_i \in L} \sum_{q=1}^{2} \sum_{R = 0}^{2} \frac{f^{(R)}_{ql}(0)}{R!}\big(\overline{\mathfrak{h}}_{\mathbf{x}_i(k)} W_{l(k)}^T\big)^R \nonumber 
\\
&= \sum_{l = 1}^M \sum_{\mathbf{x}_i \in L} \Big[\sum_{q = 1}^2 f^{(0)}_{ql}(0) + \big(\sum_{q = 1}^2 f^{(1)}_{ql}(0)\big)\overline{\mathfrak{h}}_{\mathbf{x}_i(k)} W_{l(k)}^T \nonumber
\\ 
&\textit{\ \ \ \ \ \ \ \ \ \ \ \ \ \ \ \ \ \ \ \ \ \ }+ \big(\sum_{q = 1}^2 \frac{f^{(2)}_{ql}(0)}{2!}\big)(\overline{\mathfrak{h}}_{\mathbf{x}_i(k)} W_{l(k)}^T)^2 \Big]
\label{PolyCrossEntropy}
\end{align}
where $\forall l \in [1, M]:$ $f_{1l}(z) = y_{il}\log (1 + e^{-z})$ and $f_{2l}(z) = (1-y_{il})\log (1 + e^{z})$.

To achieve $\epsilon_3$-differential privacy, we employ \textit{functional mechanism} \cite{zhang2012functional} to perturb the loss function $\widehat{\mathcal{F}}_L(\theta)$ by injecting Laplace noise into its polynomial coefficients. So, we only need to perturb $\widehat{\mathcal{F}}_L(\theta)$ just once in each training batch. To be clear, we denote $\{\phi^{(0)}_{l\mathbf{x}_i}, \phi^{(1)}_{l\mathbf{x}_i}, \phi^{(2)}_{l\mathbf{x}_i}\}$ as the coefficients, where $\phi^{(0)}_{l\mathbf{x}_i} = \sum_{q = 1}^2 f^{(0)}_{ql}(0)$ and $\phi^{(1)}_{l\mathbf{x}_i}$ and $\phi^{(2)}_{l\mathbf{x}_i}$ are coefficients at the first order and the second order of the function $\widehat{\mathcal{F}}_L(\theta)$. In fact, $\phi^{(1)}_{l\mathbf{x}_i}$ and $\phi^{(2)}_{l\mathbf{x}_i}$ will be combinations between the approximation terms $\sum_{q = 1}^2 f^{(1)}_{ql}(0)$, $\sum_{q = 1}^2 \frac{f^{(2)}_{ql}(0)}{2!}$, and $\overline{\mathfrak{h}}_{\mathbf{x}_i(k)}$.

In Alg. \ref{AdLM}, we set $\Delta_{\mathcal{F}} = M(|\overline{\mathfrak{h}}_{(k)}| + \frac{1}{4}|\overline{\mathfrak{h}}_{(k)}|^2)$ (\textit{line 17}). In essence, coefficients $\phi^{(R)}_{l\mathbf{x}_i}$ with $R \in [0, 2]$ are functions of the label $y_{il}$ only. Therefore, we can perform the perturbation by injecting Laplace noise $1/|L|Lap(\frac{\Delta_{\mathcal{F}}}{\epsilon_3})$ into $\phi^{(R)}_{l\mathbf{x}_i}$ for every training label $y_{il} \in D$ (\textit{lines 18-19}). Then, the perturbed coefficients, denoted $\overline{\phi}^{(R)}_{l\mathbf{x}_i}$ are used to construct the differentially private loss function $\overline{\mathcal{F}}_L(\theta_t)$ (\textit{line 27}) during the training process without accessing the original label $y_{il}$ again (\textit{lines 20-30}). Stochastic gradient descent and back-propagation algorithms are used to minimize the perturbed loss function $\overline{\mathcal{F}}_L(\theta_t)$.

Now, we are ready to state that the computation of $\overline{\mathcal{F}}_L(\theta_t)$ is $\epsilon_3$-differentially private, and our mechanism preserves $(\epsilon_1 + \epsilon_2 + \epsilon_3)$-differential privacy in the following lemmas.

\begin{lemma} Let $L$ and $L'$ be any two neighboring batches. Let $\widehat{\mathcal{F}}_L(\theta)$ and $\widehat{\mathcal{F}}_{L'}(\theta)$ be the loss functions on $L$ and $L'$ respectively, then we have the following inequality:
\begin{equation}
\Delta_{\mathcal{F}} = \sum_{l = 1}^M \sum_{R = 0}^2\Big\lVert \sum_{\mathbf{x}_i \in L} \phi^{(R)}_{l \mathbf{x}_i} - \sum_{\mathbf{x}'_i \in L'} \phi^{(R)}_{l \mathbf{x}'_i} \Big\rVert \leq M(|\overline{\mathfrak{h}}_{(k)}| + \frac{1}{4}|\overline{\mathfrak{h}}_{(k)}|^2) \nonumber
\end{equation}
where $|\overline{\mathfrak{h}}_{(k)}|$ is the number of hidden neurons in $\overline{\mathfrak{h}}_{(k)}$.
\label{lemma5}
\end{lemma}

\textit{Proof 5: } The proof of Lemma \ref{lemma5} is in \textbf{Appendix D}$^1$.

\begin{lemma} Algorithm \ref{AdLM} preserves $\epsilon_3$-differential privacy in the computation of $\overline{\mathcal{F}}_L(\theta_t)$ (line 27). 
\label{lemma6dd}
\end{lemma}

\textit{Proof 6: } The proof of Lemma \ref{lemma6dd} is in \textbf{Appendix E}$^1$.

\subsection{The Correctness and Characteristics of the AdLM}
The following theorem illustrates that the Alg. \ref{AdLM} preserves $\epsilon$-differential privacy, where $\epsilon = \epsilon_1+ \epsilon_2 + \epsilon_3$. 
\begin{theorem} Algorithm \ref{AdLM} preserves $\epsilon$-differential privacy, where $\epsilon = \epsilon_1 + \epsilon_2 + \epsilon_3$. 
\label{lemma6}
\end{theorem}

\textit{Proof: } At a specific training step $t \in T$, it is crystal clear that the computation of $\overline{\mathbf{h}}_{0L}$ is $\epsilon_2$-differentially private (Lemma \ref{lemma4}). Therefore, the computation of the hidden layers $\mathbf{h}_1, \ldots, \mathbf{h}_k$ and normalization layers $\overline{\mathfrak{h}}$ are differentially private. This is because they do not access any additional information from the data. At the output layer, the loss function $\overline{\mathcal{F}}_L(\theta_t)$ is $\epsilon_3$-differentially private (Lemma \ref{lemma6dd}). The computation of gradients $\frac{1}{|L|} \triangledown_{\theta_t}\overline{\mathcal{F}}_L(\theta_t)$ and descents is an optimization process, without using any additional information from the original data. Thus, ${\theta}_{t+1}$ is differentially private (line 29, Alg. \ref{AdLM}). 

This optimization process is repeated through $T$ steps without querying the original data $D$ (lines 21-30). This is done because Laplace noise is injected into input features $x_{ij}$ and coefficients ${\phi}^{(R)}_{l \mathbf{x}_i}$ as preprocessing steps (lines 3-19). Note that $\overline{x}_{ij}$ and $\overline{{\phi}}^{(R)}_{l \mathbf{x}_i}$ are associated with features $\mathbf{x}_{i}$ and the label $y_i$ respectively. $\overline{\mathbf{h}}_{0L}$ and $\overline{\mathcal{F}}_L(\theta_t)$ are computed based on $\overline{x}_{ij}$ and $\overline{{\phi}}^{(R)}_{l \mathbf{x}_i}$. As a result, the noise and privacy budget consumption will not be accumulated during the training process. 

Finally, $\overline{\mathcal{F}}_L(\theta_t)$ uses the outputs of $\overline{\mathbf{h}}_{0L}$, which essentially uses the differentially private relevances $\overline{\mathbf{R}}(D)$ as of one inputs. $\overline{\mathbf{R}}(D)$, $\overline{\mathbf{h}}_{0L}$, and $\overline{\mathcal{F}}_L(\theta_t)$ are achieved by applying $\epsilon_1, \epsilon_2,$ and $\epsilon_3$-differential privacy mechanisms. Furthermore, $\overline{\mathcal{F}}_L(\theta_t)$ and $\overline{\mathbf{h}}_{0L}$ access the same training batch $L$ at each training step. Therefore, based on the composition theorem \cite{Dwork:2009:DPR}, the total privacy budget in Alg. \ref{AdLM} must be the summation of $\epsilon_1$, $\epsilon_2$, and $\epsilon_3$. 

Consequently, Algorithm \ref{AdLM} preserves $\epsilon$-differential privacy, where $\epsilon = \epsilon_1+ \epsilon_2 + \epsilon_3$. \ \ \ \ \ \ \ \ \ \ \ \ \ \ \ \ \ \ \ \ \ \ \ \ \ \ \ \ \ \ \ \ \ \ \ \ \ \ \ \ $\blacksquare$

Note that $\Delta_{\mathbf{R}}$ and $\Delta_{\mathbf{h}_0}$ are dependent on the number of input features $d$. $\Delta_{\mathbf{R}}$ is negatively proportional to the number of tuples in $D$. The larger the size of $D$, the less noise will be injected into the private relevance $\overline{\mathbf{R}}$. $\Delta_{\mathcal{F}}$ is only dependent on the number of neurons in the last hidden layer and the output layer. In addition, $\Delta_{\mathbf{R}}, \Delta_{\mathbf{h}_0}$, and  $\Delta_{\mathcal{F}}$ do not depend on the number of training epochs. Consequently: 

\textbf{(1)} The privacy budget consumption in our model is totally independent of the number of training epochs. 

\textbf{(2)} In order to improve the model utility under differential privacy, our mechanism adaptively injects Laplace noise into features based on the contribution of each to the model output. 

\textbf{(3)} The average error incurred by our approximation approach, $\widehat{\mathcal{F}}_L(\theta)$, is bounded by a small number $M \times \frac{e^2 + 2e - 1}{e(1+e)^2}$ (Please refer to Lemma \ref{lemmaBounded} in \textbf{Appendix F}$^1$).

\textbf{(4)} The proposed mechanism can be applied to a variety of deep learning models, e.g., CNNs \cite{Lecun98}, deep auto-encoders \cite{Bengio2009}, Restricted Boltzmann Machines \cite{Smolensky1986}, convolution deep belief networks \cite{Lee:2009}, etc., as long as we can perturb the first affine transformation layer.

With these characteristics, our mechanism has a great potential to be applied in large datasets, without consuming excessive privacy budgets. In the experiment section, we will show that our mechanism leads to accurate results.

\begin{figure*}[t]
\centering
$\begin{array}{c@{\hspace{0.05in}}c@{\hspace{0.05in}}c}
\includegraphics[width=2.32in]{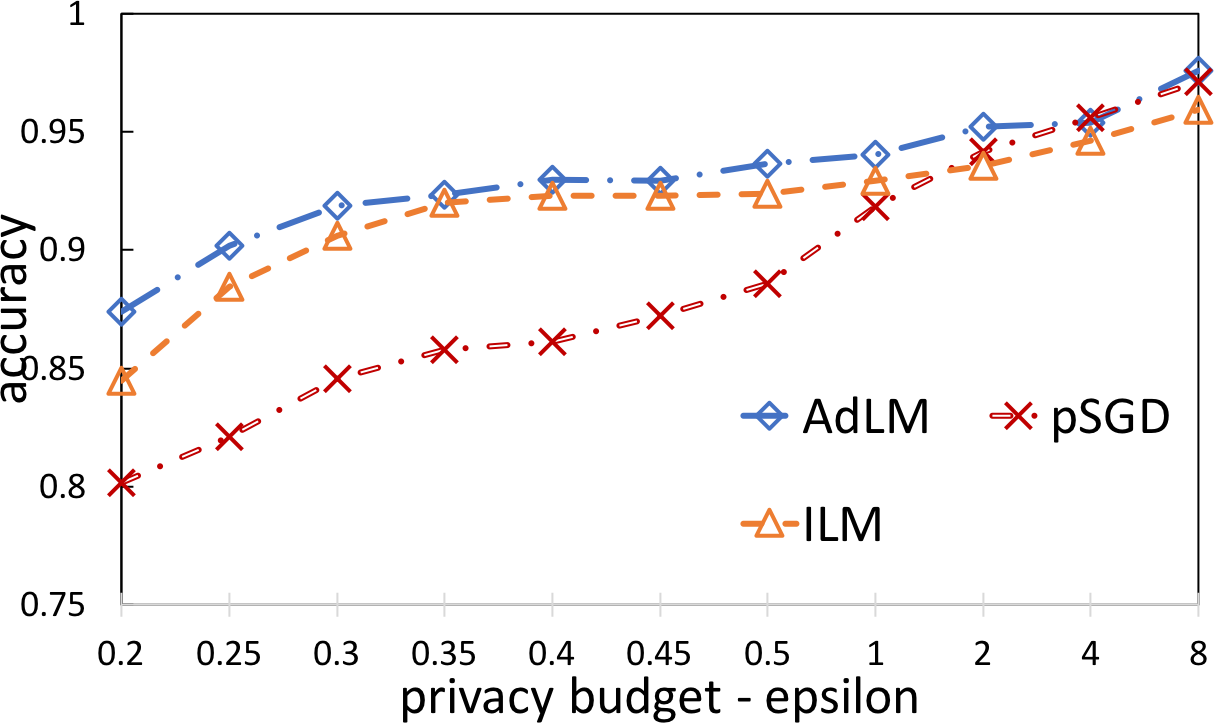} & \includegraphics[width=2.32in]{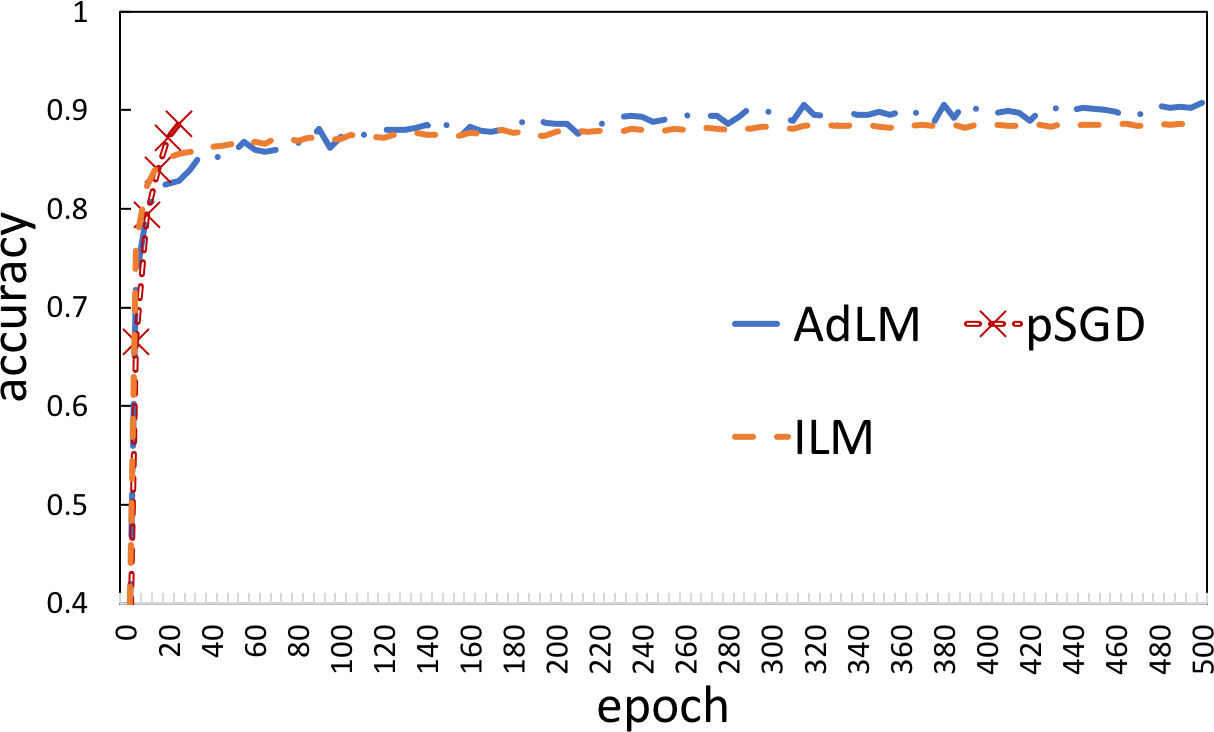} & \includegraphics[width=2.32in]{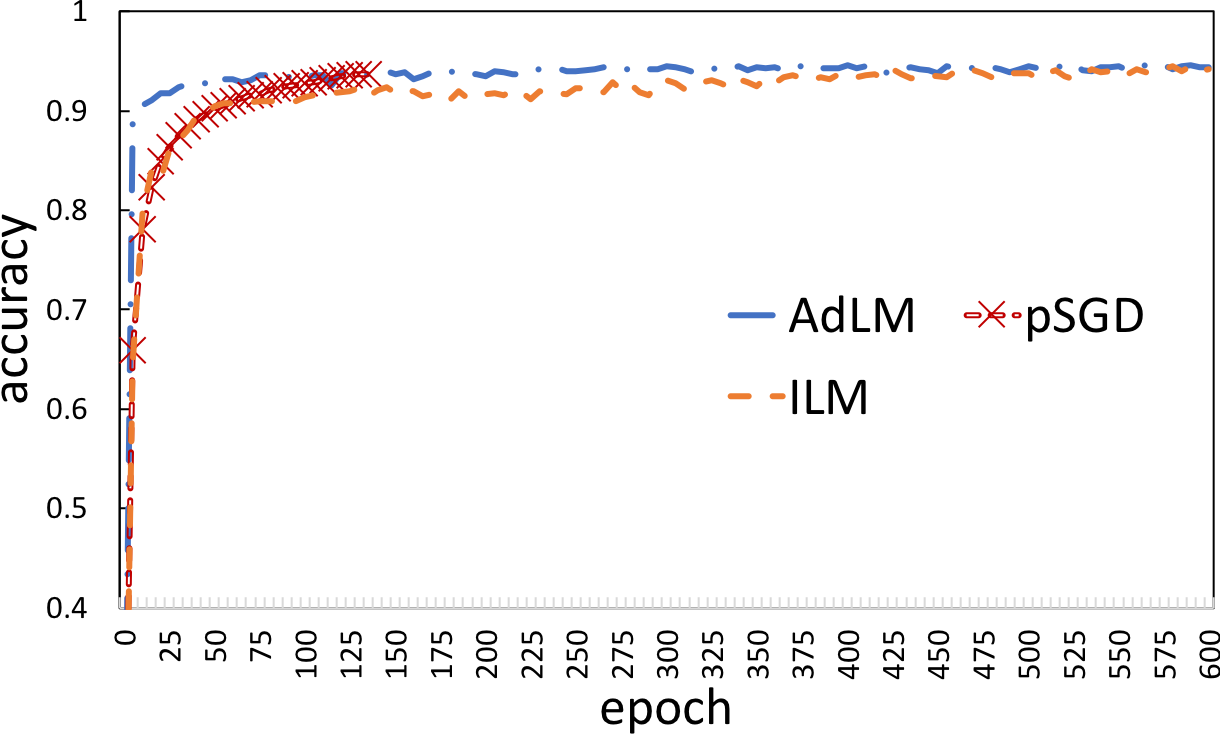} \\ [0.0cm] \mbox{(a) accuracy vs. $\epsilon$} & \mbox{(b) $\epsilon = 0.5$ (large noise)} & \mbox{(c) $\epsilon = 2.0$ (small noise)} \vspace{-5pt}
\end{array}$
\caption{Accuracy for different noise levels on the MNIST dataset.} \vspace{-5pt}
\label{MNISTnoise}
\end{figure*}

\begin{figure*}[t]
\centering
$\begin{array}{c@{\hspace{0.05in}}c@{\hspace{0.05in}}c}
\includegraphics[width=2.32in]{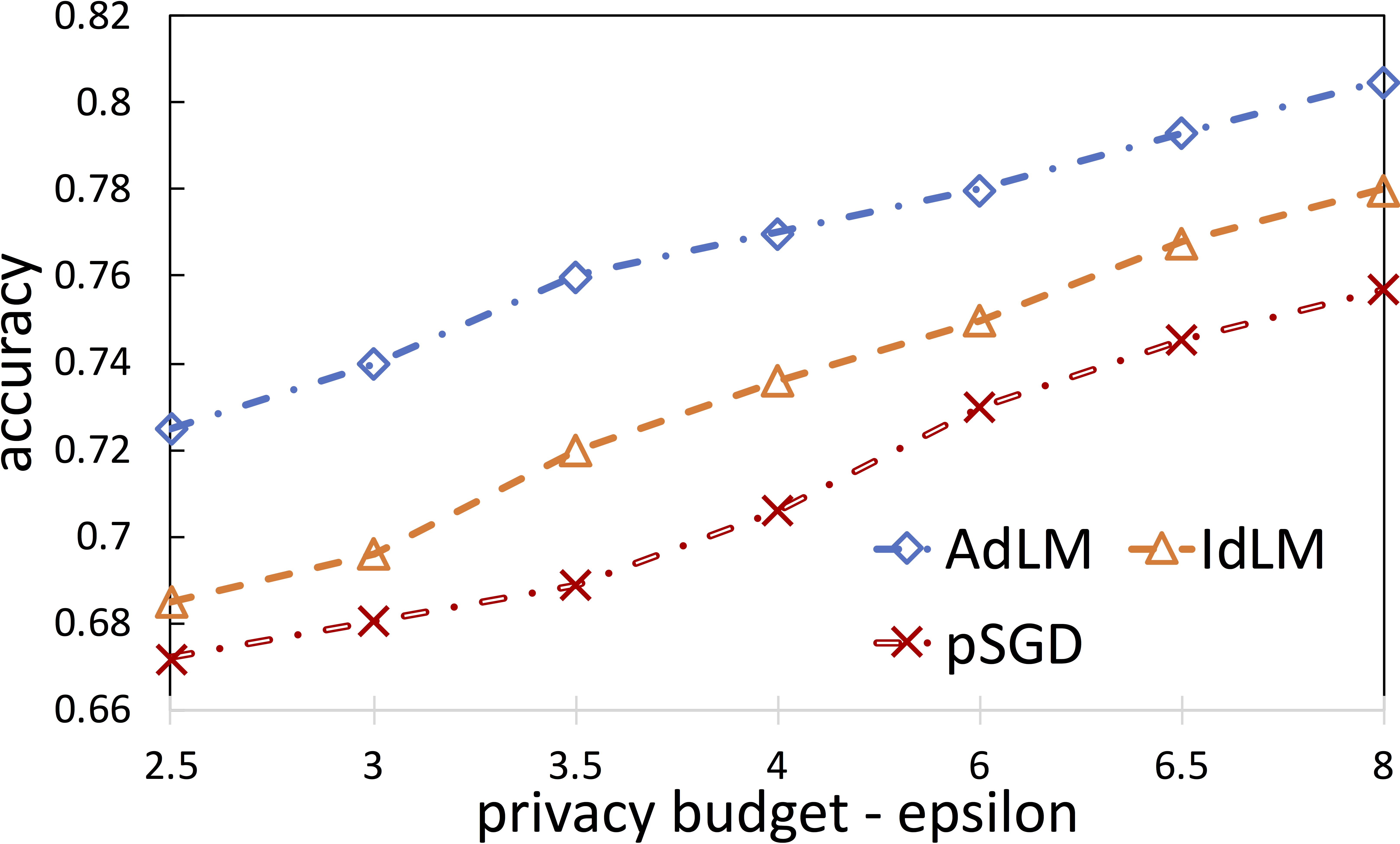} & \includegraphics[width=2.32in]{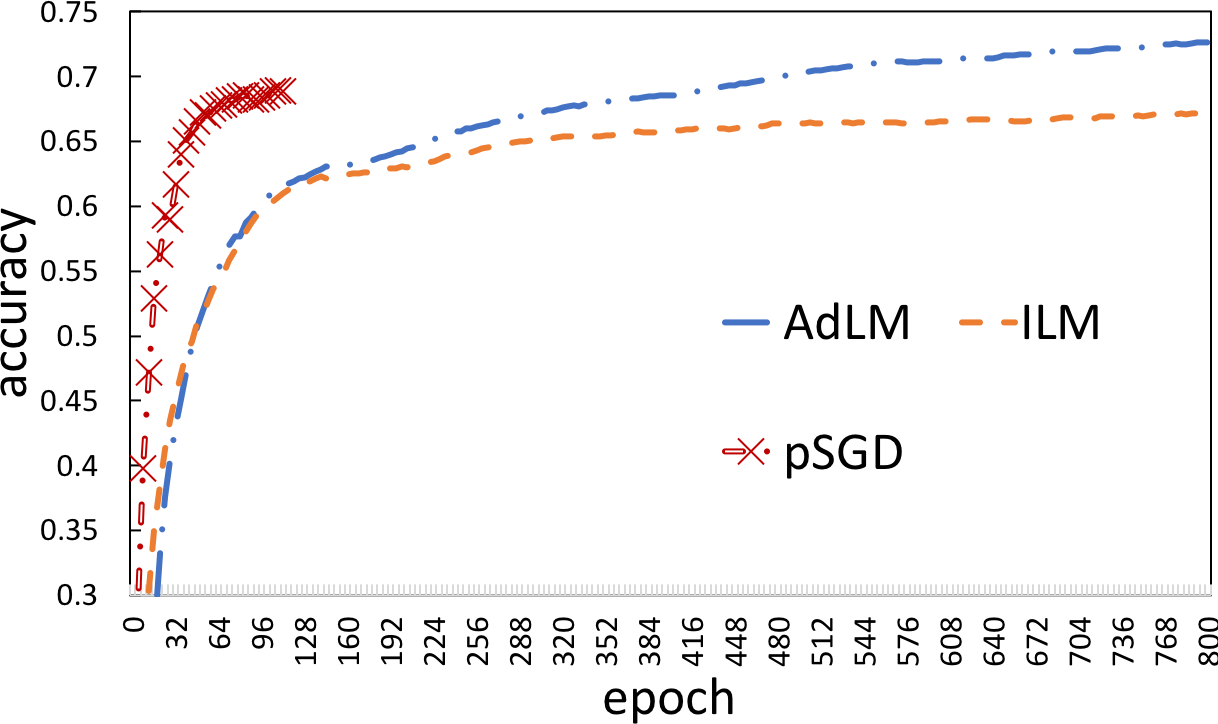} & \includegraphics[width=2.32in]{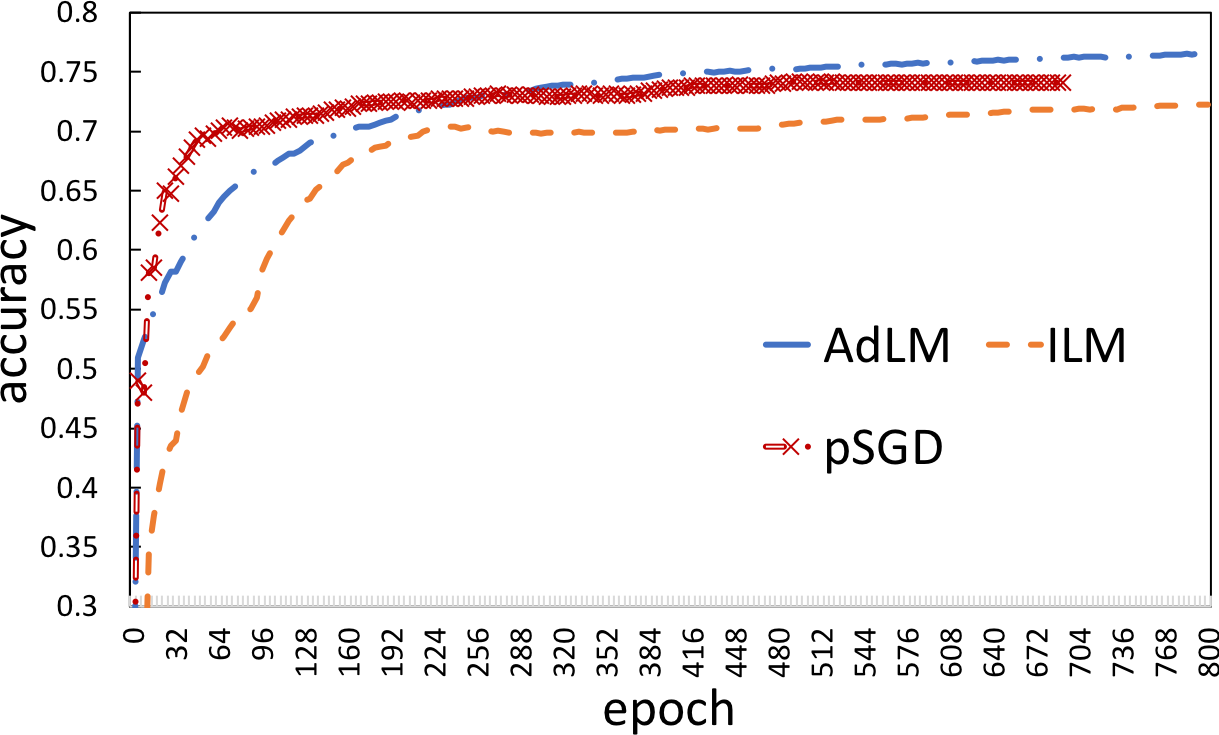} \\ [0.0cm] \mbox{(a) accuracy vs. $\epsilon$} & \mbox{(b) $\epsilon = 2.5$ (large noise)} & \mbox{(c) $\epsilon = 8.0$ (small noise)} \vspace{-5pt}
\end{array}$
\caption{Accuracy for different noise levels on the CIFAR-10 dataset.}  \vspace{-5pt}
\label{CIFARnoise}
\end{figure*}

\section{Experimental Results} 
We have carried out an extensive experiment on two well-known image datasets, MNIST and CIFAR-10. The MNIST database of handwritten digits consists of 60,000 training examples, and a test set of 10,000 examples \cite{Lecun726791}. Each example is a 28 $\times$ 28 size gray-level image. The CIFAR-10 dataset consists of color images categorized into 10 classes, such as birds, dogs, trucks, airplanes, etc. The dataset is partitioned into 50,000 training examples and 10,000 test examples \cite{krizhevsky2009learning}. 

\textbf{Competitive Models.} We compare our mechanism with the state-of-the-art differentially private stochastic gradient descent (\textbf{pSGD}) for deep learning proposed by \cite{Abadi}. CNNs are used in our experiments for both algorithms. The hyper-parameters in pSGD are set to the default values recommended by Abadi et al. \cite{Abadi}. To comprehensively examine the proposed approaches, our mechanism is implemented in two different settings: \textbf{(1)} The Adaptive Laplace Mechanism (Alg. \ref{AdLM})-based CNN with ReLUs, simply denoted \textbf{AdLM}; 
and \textbf{(2)} An \textbf{I}dentical \textbf{L}aplace \textbf{M}echanism-based CNN with ReLUs (\textbf{ILM}), in which an identical Laplace noise $\frac{1}{|L|}Lap(\frac{\Delta_{\mathbf{h}_0}}{\epsilon_2})$ is injected into each feature $x_{ij}$ to preserve $\epsilon_2$-differential privacy in the computation of the affine transformation layer $\mathbf{h}_0$. In the ILM algorithm, we do not need to use the differentially private relevances $\overline{\mathbf{R}}(D)$. 
More details about the ILM are in the \textbf{Appendix C}$^1$. \textit{The implementation of our mechanism is publicly available}\footnote{{\url{https://github.com/haiphanNJIT/PrivateDeepLearning}}}.

\subsection{MNIST Dataset}
The designs of the three models are the same on the MNIST dataset. We used two convolution layers, one with 32 features and one with 64 features. Each hidden neuron connects with a 5x5 unit patch. A fully-connected layer with 25 units and an output layer of 10 classes (i.e., 10 digits) with cross-entropy loss with LRN are used. The batch size is 1,800. This also is the structure of the pre-trained model, which is learned and used to compute the average relevances $\mathbf{R}(D)$. The experiments were conducted on a single GPU, i.e., NVIDIA GTX TITAN X, 12 GB with 3,072 CUDA cores.

Fig. \ref{MNISTnoise}a illustrates the prediction accuracy of each model as a function of the privacy budget $\epsilon$ on the MNIST dataset. It is clear that our models, i.e., AdLM and ILM, outperform the pSGD, especially when the privacy budget $\epsilon$ is small. This is a crucial result, since smaller privacy budget values enforce stronger privacy guarantees. When the privacy budget $\epsilon$ is large, e.g., $\epsilon = 2, 4, 8,$ which means small noise is injected into the model, the efficiencies of all the models are almost converged to higher prediction accuracies.

The AdLM model achieves the best performance. Given a very small privacy budget $\epsilon = 0.25$, it achieves 90.2\% in terms of prediction accuracy, compared with 88.46\% obtained by the ILM and 82.09\% obtained by the pSGD. Overall, given small values of the privacy budget $\epsilon$, i.e., $0.2 \leq \epsilon \leq 0.5$, the AdLM improves the prediction accuracy by 7.7\% on average (i.e., 91.62\%) compared with the pSGD (i.e., 83.93\%). The result is statistically significant with $p < 0.01$ (t-test).

Figs. \ref{MNISTnoise}b-c illustrate the prediction accuracy of each model vs. the number of epochs under $\epsilon = 0.5$ and $\epsilon = 2.0$ respectively. Given large noise, i.e., $\epsilon = 0.5$, the pSGD quickly achieves higher prediction accuracies (i.e., 88.59\%) after a small number of epochs, compared with other models (Fig. \ref{MNISTnoise}b). However, the pSDG can only be applied to train the model by using a limited number of epochs; specifically because the privacy budget is accumulated after every training step. Meanwhile, our mechanism is totally independent of the number of epochs in the consumption of privacy budget. Therefore, after 500 epochs, our models outperform the pSGD. The AdLM achieves the best performance, in terms of prediction accuracy: 93.66\%, whereas the ILM and the pSGD reached only 92.39\% and 88.59\%, respectively. Interestingly, given small noise, i.e., $\epsilon = 2.0$, our models achieve higher accuracies than the pSGD after a small number of epochs (Fig. \ref{MNISTnoise}c). This result illustrates the crucial benefits of being independent of the number of training epochs in preserving differential privacy in deep learning. With our mechanism, we can keep training our models without accumulating noise and privacy budget.

\subsection{CIFAR-10 Dataset}
The designs of the three models are the same on the CIFAR-10 dataset. We used three convolution layers, two with 128 features and one with 256 features. Each hidden neuron connects with a 3x3 unit patch in the first layer, and a 5x5 unit patch in other layers. One fully-connected layer with 30 neurons, and an output layer of 10 classes with a cross-entropy loss with LRN are used. The batch size is set to 7,200. This also is the structure of the pre-trained model, which is learned and used to compute the average relevances $\mathbf{R}(D)$.

Fig. \ref{CIFARnoise}a shows the prediction accuracies of each model as a function of the privacy budget $\epsilon$ on the CIFAR-10 dataset. Figs. \ref{CIFARnoise}b-c illustrate the prediction accuracy of each model vs. the number of epochs under different noise levels. Similar to the results on the MNIST dataset, the results on CIFAR-10 strengthen our observations: \textbf{(1)} Our mechanism outperforms the pSGD in terms of prediction accuracy, given both modest and large values of the privacy budget $\epsilon$ (Fig. \ref{CIFARnoise}a); and \textbf{(2)} Our mechanism has the ability to work with large-scale datasets, since it is totally independent of the number of training epochs in the consumption of privacy budget (Figs. \ref{CIFARnoise}b-c). 

In fact, the AdLM improves the prediction accuracy by 5.9\% on average (i.e., to 77\%) compared with the pSGD (i.e., 71.1\%). The result is statistically significant with $p < 0.01$ (t-test). Given large noise, i.e., $\epsilon = 2.5$, our models including the AdLM and ILM outperform the pSGD after 800 epochs (Fig. \ref{CIFARnoise}b). \vspace{-1pt}

\subsection{Adaptive Laplace Noise} \vspace{-1pt}
It is important to note that by adaptively redistributing the noise into input features based on the relevance of each to the model output, we can achieve much better prediction accuracies in both MNIST and CIFAR-10 datasets given both small and large values of privacy budget $\epsilon$. This is clearly demonstrated in Figs. \ref{MNISTnoise}a and \ref{CIFARnoise}a, since the AdLM outperforms the ILM in all cases. Overall, the AdLM improves the prediction accuracy by 2\% and 5\% on average on MNIST and CIFAR-10 datasets correspondingly, compared with the ILM. The result is statistically significant with $p < 0.05$ (t-test). Note that the ILM injected an identical amount of noise into all input features, regardless of their contributions to the model output. This is an important result, since our mechanism is the first of its kind, which can redistribute the noise injected into the deep learning model to improve the utility. In addition, the reallocation of $\epsilon_1$, $\epsilon_2$, and $\epsilon_3$ could further improve the utility. This is an open research direction in the future work. \vspace{-1pt}

\subsection{Computational Efficiency} \vspace{-1pt}
In terms of computation efficiency, there are two differences in our mechanism, compared with a regular deep neural network: \textbf{(i)} The pre-trained model; and \textbf{(ii)} The noise injection task. In practice, the pre-trained model is not necessarily identical to the differentially private network trained by our AdLM. A simple model can be used as a pre-trained model to approximate the average relevance $\mathbf{R}(D)$, as long as the pre-trained model is effective in terms of prediction accuracy even over a small training dataset. Achieving this is quite straight-forward, because: \textbf{(1)} The pre-trained model is noiseless; and \textbf{(2)} The number of training epochs used to learn a pre-trained model is small compared with the one of differentially private models. In fact, we only correspondingly need 12 and 50 extra epochs to learn the pre-trained models on MNIST and CIFAR-10 datasets. Training pre-trained models takes about 10 minutes on a single GPU, i.e., NVIDIA GTX TITAN X, 12 GB with 3,072 CUDA cores. Therefore, the model pre-training for the computation of $\mathbf{R}(D)$ is efficient.

Another difference in our mechanism is the noise injection into input attributes and coefficients of the loss function $\widehat{\mathcal{F}}$. In this task, the computations of $\Delta_{\mathbf{R}}$, $\Delta_{\mathbf{h}_0}$, $\Delta_{\mathcal{F}}$, $\overline{R}_j$, $\beta_j$, $\overline{x}$, and $\overline{\phi}$ are efficient and straight-forward, since there is not any operation such as $\arg\min$, $\arg\max$, sorting, etc. The complexity of these computations is $O\big(|D|(d + M)\big)$, which is linear to the size of the database $D$. In addition, these computations can be efficiently performed in either a serial process or a parallel process. Therefore, this task does not affect the computational efficiency of our mechanism much.

\section{Conclusions}
In this paper, we proposed a novel mechanism, called Adaptive Laplace Mechanism (AdLM), to preserve differential privacy in deep learning. Our mechanism conducts both sensitivity analysis and noise insertion on deep neural networks. It is totally independent of the number of training epochs in the consumption of privacy budget. That makes our mechanism more practical. In addition, our mechanism is the first of its kind to have the ability to redistribute the noise insertion toward the improvement of model utility in deep learning. In fact, our mechanism has the ability to intentionally add \textit{more noise} into input features which are \textit{less relevant} to the model output, and vice-versa. Different activation functions can be applied in our mechanism, as well. These distinctive characteristics guarantee the ability to apply our mechanism on large datasets in different deep learning models and in different contexts. 
Our mechanism can clearly enhance the application of differential privacy in deep learning. Rigorous experimental evaluations conducted on well-known datasets validated our theoretical results and the effectiveness of our mechanism.

\section*{Acknowledgment}
This work is supported by the NIH grant R01GM103309 to the SMASH project. Wu is also supported by NSF grant DGE-1523115 and IIS-1502273.




%
%
%


\appendix

\subsection{Proof of Lemma \ref{Lemma1}}
\begin{proof}
Assume that $D$ and $D'$ differ in the last tuple. Let $\mathbf{x}_n$ ($\mathbf{x}'_n$) be the last tuple in $D$ ($D'$). We have that
\begin{align}
\Delta_{\mathbf{R}} &= \frac{1}{|D|} \sum_{j = 1}^d \Big\lVert \sum_{\mathbf{x}_i \in D} R_{x_{ij}}(\mathbf{x}_i) - \sum_{\mathbf{x}'_i \in D'} R_{x'_{ij}}(\mathbf{x}'_i) \Big\rVert_1 \nonumber \\
&= \frac{1}{|D|} \sum_{j = 1}^d \Big\lVert R_{x_{nj}}(\mathbf{x}_n) - R_{x'_{nj}}(\mathbf{x}'_n) \Big\rVert_1 \nonumber \\
&\leq  \frac{2}{|D|} \max_{\mathbf{x}_i \in D} \sum_{j = 1}^d \lVert R_{x_{ij}}(\mathbf{x}_i) \rVert_1 \leq  \frac{2d}{|D|} \nonumber
\label{ProofLemma1}
\end{align}
Eq. \ref{GlobalSensitivity} holds.
\end{proof}

\subsection{Proof of Lemma \ref{Lemma2}}
\begin{proof}
Let $D$ and $D'$ be two neighbor databases. Without loss of generality, assume that $D$ and $D'$ differ in the last tuple $\mathbf{x}_n$ ($\mathbf{x}'_n$). $\overline{R}_j(D)$ is calculated as done in line 6, Alg. \ref{AdLM}, and $\overline{\mathcal{R}}(D) = \big\{\overline{R}_j(D)\big\}_{j \in [1, d]}$ is the output of the Alg. \ref{AdLM} (line 7). The perturbation of the relevance ${R}_j(D)$ can be rewritten as:
\begin{equation}
\overline{R}_j = \frac{1}{|D|} \sum_{\mathbf{x}_{i} \in D} R_{x_{ij}}(\mathbf{x}_i) + Lap(\frac{\Delta_{\mathbf{R}}}{\epsilon_1})
\end{equation}
Since all the input features are perturbed, we have that

\begin{align}
&\frac{Pr\big(\overline{\mathbf{R}}(D)\big)}{Pr\big(\overline{\mathbf{R}}(D')\big)} = \frac{\prod_{j = 1}^d \exp\big(\frac{\epsilon_1 \lVert \frac{1}{|D|} \sum_{\mathbf{x}_i \in D} R_j(\mathbf{x}_i) - \overline{R}_j \rVert_1}{\Delta_{\mathbf{R}}}\big)}{\prod_{j = 1}^d \exp\big(\frac{\epsilon_1 \lVert \frac{1}{|D|} \sum_{\mathbf{x}'_i \in D} R_j(\mathbf{x}'_i) - \overline{R}_j \rVert_1}{\Delta_{\mathbf{R}}}\big)} \nonumber
\\
&\leq \prod_{j = 1}^d \exp(\frac{\epsilon_1}{|D| \Delta_{\mathbf{R}}} \Big\lVert \sum_{\mathbf{x}_i \in D} R_j(\mathbf{x}_i) -  \sum_{\mathbf{x}'_i \in D'} R_j(\mathbf{x}'_i) \Big\rVert_1) \nonumber
\\
&\leq \prod_{j = 1}^d \exp(\frac{\epsilon_1}{|D| \Delta_{\mathbf{R}}} \Big\lVert R_j(\mathbf{x}_n) - R_j(\mathbf{x}'_n) \Big\rVert_1) \nonumber 
\\
&\leq \prod_{j = 1}^d \exp(\frac{\epsilon_1}{|D| \Delta_{\mathbf{R}}} 2\max_{\mathbf{x}_n \in D} \big\lVert R_j(\mathbf{x}_n) \big\rVert _1) \nonumber 
\\
&\leq \exp(\epsilon_1 \frac{2\max_{\mathbf{x}_n \in L} \sum_{j = 1}^d \lVert R_j(\mathbf{x}_n) \rVert_1}{|D| \Delta_{\mathbf{R}}}) \nonumber 
\\
&\leq \exp(\epsilon_1) \nonumber
\end{align}
Consequently, the computation of $\overline{\mathbf{R}}(D)$ preserves $\epsilon_1$-differential privacy in Alg. \ref{AdLM}. 
\end{proof}

\subsection{Identical Laplace Mechanism (ILM)}
We can add an identical noise distribution $\frac{1}{|L|}Lap(\frac{\Delta_{\mathbf{h}_0}}{\epsilon_2})$ to all input features, where $\Delta_{\mathbf{h}_0} = 2 \sum_{h \in \mathbf{h}_0} d$ (line 9, Alg. \ref{AdLM}) to preserve differential privacy in the computation of $\mathbf{h}_0$. In fact, we have that $\forall \mathbf{x}_i \in D, j \in [1, d]: \overline{x}_{ij} = x_{ij} + \frac{1}{|L|}Lap(\frac{\Delta_{\mathbf{h}_0}}{\epsilon_2})$. For each $h \in \overline{h}_{0L}$, $h$ can be re-written as:
\begin{multline}
\overline{h}_{L}(W) = \sum_{j = 1}^d \Big[\sum_{\mathbf{x}_i \in L} \big(x_{ij} + \frac{1}{|L|}Lap(\frac{\Delta_{\mathbf{h}_0}}{\epsilon_2})\big) W^T \Big] \\
+ \sum_{\mathbf{x}_i \in L}\big(b + \frac{1}{|L|}Lap(\frac{\Delta_{\mathbf{h}_0}}{\epsilon_2})\big)
\end{multline}

Let us consider the static bias $b=1$ as the $0$-th input feature and its associated parameter $W_b$, i.e., $x_{i0} = b = 1$ and $W = W_b \cup W$, we have that
\begin{align}
& \overline{h}_L(W) = \sum_{j =0}^d \Big[\sum_{\mathbf{x}_i \in L} \big(x_{ij} + \frac{1}{|L|}Lap(\frac{\Delta_{\mathbf{h}_0}}{\epsilon_2})\big) W^T \Big]
\\
& = \sum_{j = 0}^d \Big[\sum_{\mathbf{x}_i \in L} x_{ij} + Lap(\frac{\Delta_{\mathbf{h}_0}}{\epsilon_2})\Big]W^T = \sum_{j = 0}^d \overline{\phi}^h_j W^T
\end{align}
where $\overline{\phi}^h_j = \big[\sum_{\mathbf{x}_i \in L} x_{ij} + Lap(\frac{\Delta_{\mathbf{h}_0}}{\epsilon_2})\big]$.

We can see that $\overline{\phi}^h_j$ is the perturbation of the input feature $x_{ij}$ associated with the $j$-th parameter $W_j \in W$ of the hidden neuron $h$ on $L$. Since all the hidden neurons $h$ in ${\mathbf{h}_0}$ are perturbed, we have that: 
\begin{equation}
Pr\big(\overline{\mathbf{h}}_{0L}(W_0)\big) = \prod_{h \in \mathbf{h}_0} \prod_{j = 0}^d exp\big(\frac{\epsilon_2 \lVert \sum_{\mathbf{x}_i \in L}  x_{ij} -  \overline{\phi}^{h}_j\rVert}{\Delta_{\mathbf{h}_0}}\big) \nonumber 
\end{equation}

$\Delta_{\mathbf{h}_0}$ is set to $2 \sum_{h \in \mathbf{h}_0} d$ and $\overline{\mathbf{h}}_{0L}(W_0) = \{\overline{h}_L(W)\}_{h \in \mathbf{h}_0}$ is the output, we have that
\begin{align}
&\frac{Pr\big(\overline{\mathbf{h}}_{0L}(W_0)\big)}{Pr\big(\overline{\mathbf{h}}_{0L'}(W_0)\big)} = \frac{\prod_{h \in \mathbf{h}_0} \prod_{j = 0}^d \exp\big(\frac{\epsilon_2 \lVert \sum_{\mathbf{x}_i \in L}  x_{ij} -  \overline{\phi}^{h}_j\rVert_1}{\Delta_{\mathbf{h}_0}}\big)}{\prod_{h \in \mathbf{h}_0} \prod_{j = 0}^d \exp\big(\frac{\epsilon_2 \lVert \sum_{\mathbf{x}'_i \in L'} {x'}_{ij} - \overline{\phi}^{h}_j\rVert_1}{\Delta_{\mathbf{h}_0}}\big)} \nonumber
\\
&\leq \prod_{h \in \mathbf{h}_0} \prod_{j = 0}^d \exp(\frac{\epsilon_2}{\Delta_{\mathbf{h}_0}} \Big\lVert \sum_{\mathbf{x}_i \in L} x_{ij} -  \sum_{\mathbf{x}'_i \in L'} {x'}_{ij} \Big\rVert_1) \nonumber 
\\
&\leq \prod_{h \in \mathbf{h}_0} \prod_{j = 1}^d \exp(\frac{\epsilon_2}{\Delta_{\mathbf{h}_0}} \Big\lVert x_{nj} - {x'}_{nj} \Big\rVert_1) \nonumber
\\
&\leq \prod_{h \in \mathbf{h}_0} \prod_{j = 1}^d \exp(\frac{\epsilon_2}{\Delta_{\mathbf{h}_0}} 2\max_{\mathbf{x}_n \in L} \big\lVert x_{nj} \big\rVert _1) \leq \prod_{h \in \mathbf{h}_0} \prod_{j = 1}^d \exp(\frac{2\epsilon_2}{\Delta_{\mathbf{h}_0}}) \nonumber 
\\
&\leq \exp(\epsilon_2 \frac{2 \sum_{h \in \mathbf{h}_0} d}{\Delta_{\mathbf{h}_0}}) = \exp(\epsilon_2) \nonumber
\end{align}

Consequently, based on the above analysis, the computation of $\overline{\mathbf{h}}_{0L}(W_0)$ preserves $\epsilon_2$-differential privacy in Alg. \ref{AdLM} by injecting an identical Laplace noise $\frac{1}{|L|}Lap(\frac{\Delta_{\mathbf{h}_0}}{\epsilon_2})$ into all input features. In addition, given the identical Laplace noise, we do not need to use the differentially private relevance $\overline{\mathbf{R}}(D)$, since we do not need to redistribute the noise in the first affine transformation layer $\mathbf{h}_0$.

\subsection{Proof of Lemma \ref{lemma5}}
\textit{Proof 5: }
Assume that $L$ and $L'$ differ in the last tuple. Let $\mathbf{x}_n$ ($\mathbf{x}'_n$) be the last tuple in $L$ ($L'$). We have that
\begin{align}
\Delta_{\mathcal{F}} &= \sum_{l = 1}^M \sum_{R = 0}^2\Big\lVert \sum_{\mathbf{x}_i \in L} \phi^{(R)}_{l \mathbf{x}_i} - \sum_{\mathbf{x}'_i \in L'} \phi^{(R)}_{l \mathbf{x}'_i} \Big\rVert \nonumber \\
&= \sum_{l = 1}^M \sum_{R = 0}^2\big\lVert \phi^{(R)}_{l \mathbf{x}_n} - \phi^{(R)}_{l \mathbf{x}'_n} \big\rVert
\end{align}
We can show that $\phi^{(0)}_{l \mathbf{x}_n} = \sum_{q = 1}^2 f^{(0)}_{ql}(0) = y_{nl}\log 2 + (1 - y_{nl})\log 2 = \log 2$. Similarly, we can show that $\phi^{(0)}_{l\mathbf{x}'_n} = \log 2$. As a result, $\phi^{(0)}_{l \mathbf{x}_n} = \phi^{(0)}_{l \mathbf{x}'_n}$. Therefore 
\begin{align}
\Delta_{\mathcal{F}} & = \sum_{l = 1}^M \sum_{R = 0}^2\big\lVert \phi^{(R)}_{l \mathbf{x}_n} - \phi^{(R)}_{l \mathbf{x}'_n} \big\rVert = \sum_{l = 1}^M \sum_{R = 1}^2\big\lVert \phi^{(R)}_{l \mathbf{x}_n} - \phi^{(R)}_{l \mathbf{x}'_n} \big\rVert \nonumber 
\\
 & \leq \sum_{l = 1}^M \sum_{R = 1}^2 \big( \big\lVert \phi^{(R)}_{l \mathbf{x}_n} \big\rVert + \big\lVert \phi^{(R)}_{l \mathbf{x}'_n} \big\rVert \big) 
\leq 2\max_{\mathbf{x}_n} \sum_{l = 1}^M \sum_{R = 1}^2 \lVert \phi^{(R)}_{l \mathbf{x}_n} \rVert \nonumber
\\ 
 & \leq 2 \max_{\mathbf{x}_n} \Big[\sum_{l = 1}^M (\frac{1}{2} - y_{nl}) \sum_{e = 1}^{|\overline{\mathfrak{h}}_{(k)}|}\overline{\mathfrak{h}}_{e \mathbf{x}_n(k)} \nonumber 
\\
&\textit{\ \ \ \ \ \ \ \ \ \ \ \ \ \ \ \ \ \ \ \ \ \ \ \ \ } + \sum_{l = 1}^M \big(\frac{1}{8}\sum_{e, g}\overline{\mathfrak{h}}_{e \mathbf{x}_n(k)}\overline{\mathfrak{h}}_{g \mathbf{x}_n(k)}\big) \Big] \nonumber  
\\
 & \leq 2 (\frac{1}{2}M\times |\overline{\mathfrak{h}}_{(k)}| + \frac{1}{8} M \times |\overline{\mathfrak{h}}_{(k)}|^2) \nonumber 
\\
&= M(|\overline{\mathfrak{h}}_{(k)}| + \frac{1}{4}|\overline{\mathfrak{h}}_{(k)}|^2) \nonumber 
\end{align}
where $\overline{\mathfrak{h}}_{e \mathbf{x}_n(k)}$ is the state of $e$-th hidden neuron in $\overline{\mathfrak{h}}_{(k)}$.

\subsection{Proof of Lemma \ref{lemma6dd}}
\textit{Proof 6: }
Let $L$ and $L'$ be two neighbor batches. Without loss of generality, assume that $L$ and $L'$ differ in the last tuple $\mathbf{x}_n$ ($\mathbf{x}'_n$). $\Delta_\mathcal{F}$ is calculated as done in line 17, Alg. \ref{AdLM}, and $\overline{\mathcal{F}}_L(\theta_t) = \sum_{l = 1}^M \sum_{\mathbf{x}_i \in L} \sum_{R = 0}^{2} \overline{\phi}^{(R)}_{l \mathbf{x}_i}\big(\overline{\mathfrak{h}}_{\mathbf{x}_i(k)} W_{l(k)}^T\big)^R$ is the output of line 27 of the Alg. \ref{AdLM}. Note that $\overline{\mathfrak{h}}_{\mathbf{x}_i(k)}$ is the state of $\overline{\mathfrak{h}}_{(k)}$ derived from $\overline{\mathbf{h}}_{0\mathbf{x}_i}$ by navigating through the neural network. The perturbation of the coefficient $\phi^{(R)}_{l}$, denoted as $\overline{\phi}^{(R)}_{l}$, can be rewritten as:
\begin{equation}
\overline{\phi}^{(R)}_{l} = \Big[\sum_{\mathbf{x}_i \in L} \phi^{(R)}_{l \mathbf{x}_i} + Lap(\frac{\Delta_{\mathcal{F}}}{\epsilon_3})\Big]
\end{equation}

We can see that $\overline{\phi}^{(R)}_{l}$ is the perturbation of the coefficient $\phi^{(R)}_{l}$ associated with the labels $y_{il}$ in the training batch $L$. We have that
\begin{align}
&\frac{Pr\big(\overline{\mathcal{F}}_L(\theta_t)\big)}{Pr\big(\overline{\mathcal{F}}_{L'}(\theta_t)\big)} = \frac{\prod_{l = 1}^M \prod_{R = 0}^2 \exp\big(\frac{\epsilon_3 \lVert \sum_{\mathbf{x}_i \in L} \phi^{(R)}_{l \mathbf{x}_i} - \overline{\phi}^{(R)}_{l} \rVert_1}{\Delta_{\mathcal{F}}}\big)}{\prod_{l = 1}^M \prod_{R = 0}^2 \exp\big(\frac{\epsilon_3 \lVert \sum_{\mathbf{x}'_i \in L'} \phi^{(R)}_{l \mathbf{x}_i} - \overline{\phi}^{(R)}_{l} \rVert_1}{\Delta_{\mathcal{F}}}\big)} \nonumber
\\
&\leq \prod_{l = 1}^M \prod_{R = 0}^2 \exp(\frac{\epsilon_3}{\Delta_{\mathcal{F}}} \Big\lVert \sum_{\mathbf{x}_i \in L} \phi^{(R)}_{l \mathbf{x}_i} -  \sum_{\mathbf{x}'_i \in L'} \phi^{(R)}_{l \mathbf{x}'_i} \Big\rVert_1) \nonumber
\\
&\leq \prod_{l = 1}^M \prod_{R = 0}^2 \exp(\frac{\epsilon_3}{\Delta_{\mathcal{F}}} \Big\lVert \phi^{(R)}_{l \mathbf{x}_n} - \phi^{(R)}_{l \mathbf{x}'_n} \Big\rVert_1) \nonumber 
\\
&\leq \prod_{l = 1}^M \prod_{R = 0}^2 \exp(\frac{\epsilon_3}{\Delta_{\mathcal{F}}} 2\max_{\mathbf{x}_n \in L} \big\lVert \phi^{(R)}_{l \mathbf{x}_n} \big\rVert _1) \nonumber 
\\
&\leq \exp(\epsilon_3 \frac{2\max_{\mathbf{x}_n \in L} \sum_{l = 1}^M \sum_{R = 0}^2 \lVert \phi^{(R)}_{l \mathbf{x}_n} \rVert_1}{\Delta_{\mathcal{F}}}) \nonumber 
\\
&\leq \exp(\epsilon_3 \frac{M(|\overline{\mathfrak{h}}_{(k)}| + \frac{1}{4}|\overline{\mathfrak{h}}_{(k)}|^2)}{\Delta_{\mathcal{F}}}) = \exp(\epsilon_3) \nonumber
\end{align}
Consequently, the computation of $\overline{\mathcal{F}}_L(\theta_t)$ preserves $\epsilon_3$-differential privacy in Alg. \ref{AdLM}. 

\subsection{Approximation Error Bounds}
The following lemma illustrates the result of how much error our approximation approach, $\widehat{\mathcal{F}}_L(\theta)$ (Eq. \ref{PolyCrossEntropy}), incurs. The error only depends on the number of possible classification outcomes $M$. In addition, the average error of the approximations is always bounded. As in \cite{Phan0WD16,zhang2012functional}, the approximation of the loss function $\mathcal{F}_L(\theta)$ by applying Taylor Expansion without removing all polynomial terms with order larger than 2 is as follows: 
\begin{equation}
\widetilde{\mathcal{F}}_L(\theta) = \sum_{l = 1}^M \sum_{\mathbf{x}_i \in L} \sum_{q=1}^{2} \sum_{R = 0}^{\infty} \frac{f^{(R)}_{ql}(z_{ql})}{R!}\big(g_{ql}(\overline{\mathfrak{h}}_{\mathbf{x}_i(k)}, W_{l(k)}) - z_{ql}\big)^R \nonumber  
\label{Poly1}
\end{equation}
$\forall l \in \{1, \ldots, M\}$, let $f_{1l}$, $f_{2l}$, $g_{1l}$, and $g_{2l}$ be four functions defined as follows:
\begin{align}
g_{1l}(\overline{\mathfrak{h}}_{\mathbf{x}_i(k)}, W_j) &= \overline{\mathfrak{h}}_{\mathbf{x}_i(k)}W_{l(k)}^T
\\
g_{2l}(\overline{\mathfrak{h}}_{\mathbf{x}_i(k)}, W_j) &= \overline{\mathfrak{h}}_{\mathbf{x}_i(k)}W_{l(k)}^T
\\
f_{1l}(z_{1l}) &= y_{il}\log (1 + e^{-z_{1l}})
\\
f_{2l}(z_{2l}) &= (1-y_{il})\log (1 + e^{z_{2l}})
\label{4functions}
\end{align}
where $\forall q, l: z_{ql}$ is a real number.

$\forall q, l$, by setting $z_{ql} = 0$, the above equation can be simplified as: 
\begin{equation}
\widetilde{\mathcal{F}}_L(\theta) = \sum_{l = 1}^M \sum_{\mathbf{x}_i \in L} \sum_{q=1}^{2} \sum_{R = 0}^{\infty} \frac{f^{(R)}_{ql}(0)}{R!}\big(\overline{\mathfrak{h}}_{\mathbf{x}_i(k)} W_{l(k)}^T\big)^R
\label{Poly2}
\end{equation}

As in \cite{Phan0WD16}, our approximation approach works by truncating the Taylor series in Eq. \ref{Poly2} to remove all polynomial terms with order larger than 2. This leads to a new objective function in Eq. \ref{PolyCrossEntropy} with low-order polynomials as follows: 
\begin{align}
&\widehat{\mathcal{F}}_L(\theta) = \sum_{l = 1}^M \sum_{\mathbf{x}_i \in L} \sum_{q=1}^{2} \sum_{R = 0}^{2} \frac{f^{(R)}_{ql}(0)}{R!}\big(\overline{\mathfrak{h}}_{\mathbf{x}_i(k)} W_{l(k)}^T\big)^R \nonumber 
\\
&= \sum_{l = 1}^M \sum_{\mathbf{x}_i \in L} \Big[\sum_{q = 1}^2 f^{(0)}_{ql}(0) + \big(\sum_{q = 1}^2 f^{(1)}_{ql}(0)\big)\overline{\mathfrak{h}}_{\mathbf{x}_i(k)} W_{l(k)}^T \nonumber 
\\
& \textit{\ \ \ \ \ \ \ \ \ \ \ \ \ \ \ \ \ \ \ \ \ } + \big(\sum_{q = 1}^2 \frac{f^{(2)}_{ql}(0)}{2!}\big)(\overline{\mathfrak{h}}_{\mathbf{x}_i(k)} W_{l(k)}^T)^2 \Big] \nonumber
\end{align}

We are now ready to state the following lemma to show the approximation error bound of our approach.
\begin{lemma} Given two polynomial functions $\widetilde{\mathcal{F}}_L(\theta)$ (Eq. \ref{Poly2}) and $\widehat{\mathcal{F}}_L(\theta)$ (Eq. \ref{PolyCrossEntropy}), the average error of the approximation is always bounded as follows: 
\begin{equation} 
|\widetilde{\mathcal{F}}_L(\widehat{\theta}) - \widetilde{\mathcal{F}}_L(\widetilde{\theta})| \leq M \times \frac{e^2 + 2e - 1}{e(1+e)^2}
\label{bounded}
\end{equation} 
where $\widetilde{\theta} = \arg \min_{\theta} \widetilde{\mathcal{F}}_L(\theta)$ and $\widehat{\theta} = \arg\min_{\theta} \widehat{\mathcal{F}}_L(\theta)$.
\label{lemmaBounded}
\end{lemma}

\textit{Proof 7: }
Let $\widetilde{\theta} = \arg \min_\theta \widetilde{\mathcal{F}}_L(\theta)$ and $\widehat{\theta} = \arg \min_\theta \widehat{\mathcal{F}}_L(\theta)$, $U = \max_\theta \big(\widetilde{\mathcal{F}}_L(\theta) - \widehat{\mathcal{F}}_L(\theta)\big)$ and $S = \min_\theta \big(\widetilde{\mathcal{F}}_L(\theta) - \widehat{\mathcal{F}}_L(\theta)\big)$.  We have that $U \geq \widetilde{\mathcal{F}}_L(\widehat{\theta}) - \widehat{\mathcal{F}}_L(\widehat{\theta})$ and $\forall \theta^*: S \leq \widetilde{\mathcal{F}}_L(\theta^*) - \widehat{\mathcal{F}}_L(\theta^*)$. Therefore, we have
\begin{align}
&\widetilde{\mathcal{F}}_L(\widehat{\theta}) - \widehat{\mathcal{F}}_L(\widehat{\theta}) - \widetilde{\mathcal{F}}_L(\theta^*) + \widehat{\mathcal{F}}_L(\theta^*) \leq U - S \nonumber 
\\
&\Leftrightarrow \widetilde{\mathcal{F}}_L(\widehat{\theta}) - \widetilde{\mathcal{F}}_L(\theta^*) \leq U - S + \big(\widehat{\mathcal{F}}_L(\widehat{\theta}) - \widehat{\mathcal{F}}_L(\theta^*)\big) \nonumber
\end{align}

In addition, $\widehat{\mathcal{F}}_L(\widehat{\theta}) - \widehat{\mathcal{F}}_L(\theta^*) \leq 0$, so $\widetilde{\mathcal{F}}_L( \widehat{\theta}) - \widetilde{\mathcal{F}}_L(\theta^*) \leq U - S$. If $U \geq 0$ and $S \leq 0$ then we have:
\begin{equation}
|\widetilde{\mathcal{F}}_L(\widehat{\theta}) - \widetilde{\mathcal{F}}_L(\theta^*)| \leq U - S 
\label{proof1}
\end{equation}

Eq. \ref{proof1} holds for every $\theta^*$. Therefore, it still holds for $\widetilde{\theta}$. Eq. \ref{proof1} shows that the error incurred by truncating the Taylor series approximate function depends on the maximum and minimum values of $\widetilde{\mathcal{F}}_L(\theta) - \widehat{\mathcal{F}}_L(\theta)$. To quantify the magnitude of the error, we first rewrite $\widetilde{\mathcal{F}}_L(\theta) - \widehat{\mathcal{F}}_L(\theta)$ as: 
\begin{align}
&\widetilde{\mathcal{F}}_L(\theta) - \widehat{\mathcal{F}}_L(\theta) = \sum_{l=1}^M \Big[\widetilde{\mathcal{F}}_L(W_{l(k)}) - \widehat{\mathcal{F}}_L(W_{l(k)})\Big] \nonumber 
\\
&= \sum_{l = 1}^M \Big[\sum_{\mathbf{x}_i \in L} \sum_{q=1}^{2} \sum_{R = 3}^{\infty} \frac{f^{(R)}_{ql}(z_{ql})}{R!}\big(g_{ql}(\overline{\mathfrak{h}}_{\mathbf{x}_i(k)}, W_{l(k)}) - z_{ql}\big)^R \Big] \nonumber
\end{align}

To derive the minimum and maximum values of the function above, we look into the remainder of the Taylor Expansion for each $l$.
Let $z_l \in [z_{ql} - 1, z_{ql} + 1]$. According to the well-known result \cite{Apostol}, $\frac{1}{|D|}\big(\widetilde{\mathcal{F}}_L(W_{l(k)}) - \widehat{\mathcal{F}}_L(W_{l(k)})\big)$ must be in the interval $\Big[\sum_q \frac{\min_{z_l} f^{(3)}_{ql}(z_l)(z_l - z_{ql})^3}{6}, \sum_l \frac{\max_{z_l} f^{(3)}_{ql}(z_l)(z_l - z_{ql})^3}{6}\Big]$. 

If $\sum_q \frac{\max_{z_l} f^{(3)}_{ql}(z_l)(z_l - z_{ql})^3}{6} \geq 0$ and $\sum_q \frac{\min_{z_l} f^{(3)}_{ql}(z_l)(z_l - z_{ql})^3}{6} \leq 0$, then we have that: 
\begin{multline}
\Big|\frac{1}{|L|}\Big[\widetilde{\mathcal{F}}_L(\theta) - \widehat{\mathcal{F}}_L(\theta)\Big] \Big| \\
\leq \sum_{l =1}^M \sum_q \frac{\max_{z_l} f^{(3)}_{ql}(z_l)(z_l - z_{ql})^3 - \min_{z_l} f^{(3)}_{ql}(z_l)(z_l - z_{ql})^3}{6}
\end{multline} 

This analysis applies to the case of the cross-entropy error-based loss function as follows. First, for the functions $f_{1l}(z_{1l}) = y_{il}\log (1 + e^{-z_{1l}})$ and $f_{2l}(z_{2l}) = (1-y_{il})\log (1 + e^{z_{2l}})$, we have 
\begin{align}
f_{1l}^{(3)}(z_{1l}) &= \frac{2y_{il}e^{z_{1l}}}{(1 + e^{z_{1l}})^3} \nonumber
\\
f_{2l}^{(3)}(z_{2l}) &= (1-y_{il})\frac{e^{-z_{2l}}(e^{-z_{2l}} -1)}{(1+ e^{-z_{2l}})^3} \nonumber
\end{align}

It can be verified that $\arg \min_{z_{1l}}f_{1l}^{(3)}(z_{1l}) = \frac{-2e}{(1+e)^3} < 0$, $\arg \max_{z_{1l}}f_{1l}^{(3)}(z_{1l}) =\frac{2e}{(1+e)^3} > 0$, $\arg \min_{z_{2l}}f_{2l}^{(3)}(z_{2l}) = \frac{1-e}{e(1+e)^3} < 0$, and $\arg \max_{z_{2l}}f_{2l}^{(3)}(z_{2l}) = \frac{e(e-1)}{(1+e)^3} > 0$. Thus, the average error of the approximation is at most 
\begin{multline}
\big|\widetilde{\mathcal{F}}_L(\widehat{\theta}) - \widetilde{\mathcal{F}}_L(\widetilde{\theta})\big| \leq M \times \Big[\big( \frac{2e}{(1+e)^3} - \frac{-2e}{(1+e)^3}\big) 
\\
+ \big(\frac{e(e-1)}{(1+e)^3} - \frac{1-e}{e(1+e)^3}\big)\Big]= M \times \frac{e^2 + 2e - 1}{e(1+e)^2} \nonumber
\end{multline}

Therefore, Eq. \ref{bounded} holds.

\subsection{Corrections of the Paper} 
The differences between this correction and the first submission of our paper can be summarized as follows: 
\begin{itemize}
\item [1.] There was a mistake in terms of model configurations reported in our original version. The number of layers, hidden neurons, unit patches, and batch sizes are updated in this correction version. The experimental results of our algorithms and the differentially private Stochastic Gradient Descent algorithm (\textbf{pSGD}) \cite{Abadi} are updated as well. In fact, the code release of the pSGD algorithm\footnote{\scriptsize{\url{https://github.com/tensorflow/models/tree/master/research/differential_privacy}}} is used in this version. The pSGD algorithm is significantly improved in terms of accuracy, and the computation of the privacy budget $\epsilon$ is also more accurate. 

\item [2.] We replace the equations of coefficients $\{\phi^{(0)}_{l\mathbf{x}_i},$ $\phi^{(1)}_{l\mathbf{x}_i},$ $\phi^{(2)}_{l\mathbf{x}_i}\}$, i.e., after the Eq. 25, with a more detailed explanation, since it may cause some misunderstanding to the readers in terms of identifying coefficients of the Taylor Expansion in Eq. 25. To be clear, we denote $\{\phi^{(0)}_{l\mathbf{x}_i}, \phi^{(1)}_{l\mathbf{x}_i}, \phi^{(2)}_{l\mathbf{x}_i}\}$ as the coefficients, where $\phi^{(0)}_{l\mathbf{x}_i} = \sum_{q = 1}^2 f^{(0)}_{ql}(0)$ and $\phi^{(1)}_{l\mathbf{x}_i}$ and $\phi^{(2)}_{l\mathbf{x}_i}$ are coefficients at the first order and the second order of the function $\widehat{\mathcal{F}}_L(\theta)$. $\phi^{(1)}_{l\mathbf{x}_i}$ and $\phi^{(2)}_{l\mathbf{x}_i}$ are combinations between the approximation terms $\sum_{q = 1}^2 f^{(1)}_{ql}(0)$, $\sum_{q = 1}^2 \frac{f^{(2)}_{ql}(0)}{2!}$, and $\overline{\mathfrak{h}}_{\mathbf{x}_i(k)}$.

\end{itemize}

\end{document}